\documentclass[aps,prd,superscriptaddress,nofootinbib,tighten,preprint]{revtex4}
\usepackage[utf8]{inputenc}
\usepackage[english]{babel}
\usepackage{amsmath}
\usepackage{amsmath}
\usepackage{subfigure}
\usepackage{cancel}
\usepackage{amsfonts}
\usepackage{amssymb}
\usepackage{mathtools}
\usepackage{graphicx}
\usepackage[utf8]{inputenc}
\usepackage[T1]{fontenc}
\usepackage{xcolor}
\newcommand{\be}{\begin{equation}}
\newcommand{\ee}{\end{equation}}
\newcommand{\bea}{\begin{eqnarray}}
\newcommand{\eea}{\end{eqnarray}}

\newcommand{\zmt}{Z_{BL}}

\catcode`@=12

\begin{document}

\title{Evading Dark Matter Bounds through NLSP-Assisted Freeze-Out with Long-Lived Signatures}

\author{Sarif Khan}
\email{sarifkhan@cau.ac.kr}
\affiliation{Department of Physics, Chung-Ang University, Seoul 06974, Korea.}

\section*{}

\vspace{1cm}

\begin{abstract} 

In this work, we explore a conversion-driven freeze-out scenario, where the next-to-lightest stable particle (NLSP) sets the dark matter (DM) abundance through the process 
``NLSP SM $\leftrightarrow$ DM SM". Although DM is produced via a freeze-out mechanism, its interaction strength with the visible sector can range from weak to feeble couplings.
This results in a vast, largely unexplored parameter space that evades current direct, indirect, and collider bounds, while remaining testable in the near future.
We study this mechanism in the context of an alternative $U(1)_{B-L}$ model, where four
chiral fermions are required to cancel gauge anomalies, unlike the usual case with three 
right-handed neutrinos. The observed relic abundance is 
successfully reproduced within this framework. The viable parameter space can be probed 
by future direct detection experiments, while remaining inaccessible to indirect searches.
Our results show that the DM relic density is highly sensitive to the NLSP-SM interaction strength and the mass difference between the NLSP and DM, but not to the DM-SM direct interaction. For certain parameter choices, the NLSP decays to DM via two or three body processes involving an extra gauge boson and SM particles, leading to long-lived decays outside the CMS or ATLAS detectors at the LHC. In contrast, if the decay proceeds via a CP-odd Higgs, it occurs promptly within the detector.
We investigate prospects for detecting such long-lived NLSPs at the proposed MATHUSLA detector, with similar expectations for the ongoing FASER experiment. Finally, we find that choosing arbitrarily small values of the gauge coupling or BSM fermionic mixing angle can violate successful BBN predictions.

\end{abstract}
\maketitle

\section{Introduction}
\label{Intro}
The Standard Model (SM) of particle physics has remained a remarkably successful theory, 
requiring no extensions to its gauge group or particle content, despite extensive 
experimental searches for physics beyond the Standard Model (BSM). 
However, there are a few BSM phenomena which
cannot be addressed within the framework of the SM. The most significant drawback 
in this regard is the presence of DM has been confirmed by many evidences,
starting from the flatness of galactic rotation curves \cite{Sofue:2000jx}, the observation of the Bullet Cluster with the non-coincidence of the centre of mass of visible and total matter \cite{Clowe:2003tk, Harvey:2015hha}, 
as well as the precise measurement of DM abundance required to explain the 
CMB anisotropy \cite{Planck:2015fie, Planck:2018vyg}. 
In the SM, we do not have any suitable candidate that satisfies all the properties 
of DM as listed in Ref. \cite{Taoso:2007qk}. Therefore, it is strongly motivated to 
add an additional degree of freedom as a DM candidate by extending either 
the gauge sector and/or the particle spectrum. Moreover, another BSM problem 
is the existence of neutrino mass, confirmed by a number of neutrino 
oscillation experiments \cite{Super-Kamiokande:1998kpq, SNO:2002tuh, KamLAND:2002uet}. 
To accommodate the neutrino mass, we also need to extend the SM, and this has been studied in the context of the present particle framework in 
Refs. \cite{Patra:2016ofq, Biswas:2018yus, Khan:2025yko}, so neutrino mass will 
not be discussed in the present work.
In this work, we mainly focus on DM production using the conversion-driven freeze-out (CDFO) mechanism \cite{Garny:2017rxs, DAgnolo:2017dbv, Garny:2018icg, DiazSaez:2024nrq, Zhang:2024sox}, which will be discussed in detail later on.

DM has been extensively studied with weak-scale coupling strength, yet no detection has occurred so far. 
In particular, direct detection (DD) experiments have explored a vast parameter space, and in the near future, the exclusion limits are expected to reach the neutrino floor. 
The present stringent limit comes from LUX-ZEPLIN \cite{LZ:2022lsv, LZ:2024zvo}, 
which has ruled out 
scenarios where DM has direct coupling with the visible sector, as in such cases, achieving the correct DM relic density while evading DD bounds is not possible simultaneously.
As an example, a direct interaction between scalar DM and the SM Higgs requires the interaction strength to be below $\mathcal{O}(10^{-3})$
  to satisfy the current LUX-ZEPLIN DD bound, whereas achieving the correct DM relic abundance requires a value $\mathcal{O}(0.1)$ which differs significantly\footnote{Although the situation changes near the resonance region.}.
In indirect detection (ID) experiments as well, the exclusion limit from Fermi-LAT \cite{MAGIC:2016xys} 
has reached the thermal cross-section limit for DM, which accounts for DM correct relic density. 
At colliders, there have been extensive DM searches, particularly 
in the context of supersymmetry, which has explored a large parameter space
without any positive signal for a DM candidate 
\cite{ATLAS:2024qmx, Lee:2025mzh, Constantin:2025mex}. 
Moreover, precise measurements of Higgs observables have constrained the Higgs branching ratios to invisible degrees of freedom to a narrow range and the recent data indicates $Br(h_{1} \rightarrow inv) \leq 0.16$ \cite{CMS:2018yfx, CMS:2021far, CMS:2020ulv}. 
Therefore, it is possible that our dark matter lies in a different parameter regime with a different production mechanism. It is thus timely to explore alternative scenarios that can explain the current status of dark matter. The present work contributes to this direction.
 
There have been many different mechanisms for DM production, 
starting from the freeze-out mechanism \cite{Gondolo:1990dk}, 
which results in a weakly interacting 
massive particle (WIMP), or the freeze-in mechanism \cite{Hall:2009bx}, 
which results in 
a feebly interacting massive particle (FIMP). 
As mentioned, WIMP DM with direct interaction to the SM is in conflict with DD data,
but there are many ways to evade the DD bounds. One of the popular ways to evade 
these bounds is through the introduction of mixing between DM and the visible sector, 
or DM annihilation into other dark sector particles.
Moreover, studying a multicomponent DM scenario is also very effective in 
evading DD and other experimental bounds \cite{Khan:2025yko, Covi:2025erx}. 
On the other hand, 
the FIMP DM candidate, which has feeble coupling with the visible sector, 
is very hard to detect in different experiments. Although it explains the non-detection 
of DM so far and successfully generates the correct DM abundance, but it is not very appealing in the context of detection prospects. 
However, recently, there have been attempts to detect FIMP DM in the context of low reheat temperature, which opens up DM production via the freeze-in 
mechanism at stronger coupling \cite{Arcadi:2024wwg, Silva-Malpartida:2023yks,
Lee:2024wes,Khan:2025keb}.
In the present work, we have followed the CDFO mechanism, where we 
need a next-to-lightest stable particle (NLSP) and a DM candidate that 
interacts with the Higgs bosons or gauge bosons. In this setup, the NLSP 
undergoes freeze-out via ``NLSP NLSP $\leftrightarrow$ SM SM'', and 
the DM undergoes freeze-out via ``NLSP SM $\leftrightarrow$ DM SM'' or both NLSP and DM freeze out 
through the latter process. Therefore, we 
can freely choose the DM interaction with the visible sector and 
evade different experimental bounds.
Thus, DM production via the CDFO mechanism opens up a large parameter space that remains unexplored by current and future direct, indirect, or collider experiments. 
In Refs. \cite{Garny:2017rxs, DAgnolo:2017dbv, Garny:2018icg, DiazSaez:2024nrq, Zhang:2024sox}, CDFO has been explored earlier in different kind of
particle setup and our work is in the context of 
well motivated $U(1)_{B-L}$ ultraviolet extension of the SM.

In this work, we have extended the SM by an additional gauge symmetry 
$U(1)_{B-L}$ where $B$ is baryon number and $L$ is lepton number, 
and to make the model anomaly-free, we have added four chiral 
fermions \cite{Patra:2016ofq, Biswas:2018yus, Khan:2025yko}, in contrast to the more popular addition of three 
right-handed neutrinos. To generate mass for the chiral fermions, 
we have also introduced two singlet scalars, and once they acquire spontaneous vevs, the extra fermions obtain their masses.
Out of the four chiral fermions, we can construct two Dirac fermions, 
one of which acts as the NLSP (defined as $\psi_2$) and the other one as the DM candidate ($\psi_1$). 
In this same particle setup, WIMP DM \cite{Nanda:2017bmi}, 
FIMP DM \cite{Khan:2025yko}, and multi-component 
DM \cite{Biswas:2018yus, Khan:2025yko} scenarios have been studied in the literature. 
In the present work, we study the CDFO mechanism, where we consider $\psi_1$ as the DM and $\psi_2$ as the NLSP, which decays to DM after freeze-out, similar to the superWIMP mechanism \cite{Covi:1999ty}. 
We find that the DM abundance is controlled by 
the NLSP abundance and the mass difference between the NLSP and DM. Since the 
DM abundance is primarily set by the NLSP abundance, we can freely choose the 
DM interaction with the visible sector. This opens up a large parameter space 
that remains unexplored by current experiments.
We find that, for suitable choices of model parameters, our DD cross section lies just below the LUX-ZEPLIN limit but within the reach of the Darwin experiment. We also 
find that the exploration of DM in ID experiments is unpromising due to the double suppression,
one from p-wave annihilation due to the fermionic nature of DM and 
the other one from the small interaction between DM and the 
visible sector.
We have also demonstrated the detection prospects of the NLSP at the 
MATHUSLA detector \cite{Curtin:2018mvb} 
(with similar prospects at FASER \cite{FASER:2022hcn} expected). 
Depending on the choice 
of parameter values, $\psi_2$ can decay dominantly via the two-body channels 
$\psi_2 \rightarrow \psi_1 A$ or $\psi_2 \rightarrow \psi_1 Z_{BL}$, or 
via the three-body decay $\psi_2 \rightarrow \psi_1 f \bar{f}$.
Moreover, we find that certain parameter regions scanned during our analysis may contradict BBN and CMB physics, and we have identified and presented 
those parameter spaces in the decay length vs. abundance plane.

The rest of the paper is organised as follows. In Section \ref{model}, we describe the model part of our work. Section \ref{constraints} discusses the relevant constraints considered in our study. We present our results for DM production and the allowed parameter range in Section \ref{result}. In Sections (\ref{mathusla-section}, 
\ref{bbn-section}), we highlight the prospects associated with long-lived particles at MATHUSLA and the potential issues related to BBN and CMB. Finally, in Section \ref{conclusion}, we conclude our work, which is followed by an Appendix containing analytical expressions.

\section{The model}
\label{model}
The particle content considered in this work is presented in Tables \ref{tab1} and \ref{tab2}. We closely follow Ref.~\cite{Khan:2025yko}, to which interested readers are referred for a detailed discussion.

\begin{center}
\begin{table}[h!]
\begin{tabular}{||c|c|c|c||}
\hline
\hline
\begin{tabular}{c}
    Gauge\\
    Group\\ 
    \hline
    
    ${\rm SU(2)}_{\rm L}$\\  
    \hline
    ${\rm U(1)}_{\rm Y}$\\ 
    \hline
    $U(1)_{B-L}$\\ 
\end{tabular}
&

\begin{tabular}{c|c|c}
    \multicolumn{3}{c}{Quarks}\\ 
    \hline
    $Q_{L}^{i}=(u_{L}^{i},d_{L}^{i})^{T}$&$u_{R}^{i}$&$d_{R}^{i}$\\ 
    \hline
    $2$&$1$&$1$\\ 
    \hline
    $1/6$&$2/3$&$-1/3$\\ 
    \hline
    $1/3$&$1/3$&$1/3$\\ 
\end{tabular}
&
\begin{tabular}{c|c}
    \multicolumn{2}{c}{Leptons}\\
    \hline
    $L_{L}^{i}=(\nu_{L}^{i},e_{L}^{i})^{T}$ & $e_{R}^{i}$\\
    \hline
    $2$&$1$\\
    \hline
    $-1/2$&$-1$\\
    \hline
    $-1$&$-1$\\
\end{tabular}
&
\begin{tabular}{c}
    \multicolumn{1}{c}{Higgs doublet}\\
    \hline
    $\phi_{h}$\\
    \hline
    $2$\\
    \hline
    $1/2$\\
    \hline
    $0$\\
\end{tabular}\\
\hline
\hline
\end{tabular}
\caption{Charge assignments of SM particles under the additional 
$U(1)_{B-L}$ gauge symmetry, in addition to the SM gauge groups.}
\label{tab1}
\end{table}
\end{center}

 \begin{center}
\begin{table}[h!]
\begin{tabular}{||c|c|c||}
\hline
\hline
\begin{tabular}{c}
    Gauge\\
    Group\\ 
    \hline
    
    ${\rm SU(2)}_{\rm L}$\\  
    \hline
    ${\rm U(1)}_{\rm Y}$\\ 
    \hline
    $U(1)_{B-L}$\\ 
\end{tabular}
&

\begin{tabular}{c|c|c|c}
    \multicolumn{4}{c}{Extra fermions}\\ 
    \hline
    $\xi_{1L}$&$\xi_{2L}$&$\chi_{1R}$ &$\chi_{2R}$ \\ 
    \hline
    $1$&$1$&$1$&$1$\\ 
    \hline
    $0$&$0$&$0$&$0$\\ 
    \hline
    $4/3$&$1/3$&$-2/3$ &$-2/3$\\ 
\end{tabular}
&
\begin{tabular}{c|c}
    \multicolumn{2}{c}{Extra scalars}\\
    \hline
    $\phi_{1}$ & $\phi_{2}$\\
    \hline
    $1$ & $1$ \\
    \hline
    $0$ & $0$\\
    \hline
    $1$ & $2$\\
\end{tabular}\\
\hline
\hline
\end{tabular}
\caption{Charges of the beyond SM particles under the SM and $U(1)_{B-L}$ 
gauge groups.}
\label{tab2}
\end{table}
\end{center}

The complete Lagrangian for the particle content shown in 
Tables \ref{tab1} and \ref{tab2} takes the following form,
\begin{eqnarray}
\mathcal{L}& = &\mathcal{L}_{SM} + 
\sum_{i =1, 2} \left( D_{\mu} \phi_{i} \right)^{\dagger} \left( D_{\mu} \phi_{i} \right) 
+ \mathcal{L}^{BSM}_{BL}
 - \mathcal{V}\left(\phi_{h},\phi_{1}, \phi_{2} \right)\,.   
\end{eqnarray}
where $\mathcal{L}_{SM}$ is the SM Lagrangian except the Higgs potential,
second term represents the kinetic terms for $\phi_{1,2}$.
The third term
$\mathcal{L}^{BSM}_{BL}$ is the term associated with the BSM 
fermions as follows,
\begin{eqnarray}
\mathcal{L}^{Kin}_{BL} &=& \sum_{X = \xi_{1L}, \xi_{2L}, \chi_{1R}, \chi_{2R}} \bar{X} i \cancel{D} X + \alpha_{1} \bar \xi_{1L} \chi_{1R} \phi_{2} 
+ \alpha_{2} \bar \xi_{2L} \chi_{2R} \phi_{1} 
+ \beta_{1} \bar \xi_{2L} \chi_{1R} \phi_{1} \nonumber \\ &+& 
 \beta_{2} \bar \xi_{1L} \chi_{2R} \phi_{2} + {\it h.c.}
\label{Yukawa-exotic-fermion}
\end{eqnarray}
where $\cancel{D} X \equiv \gamma^{\mu} D_{\mu} X$ and the covariant derivative is $D_{\mu} X = \partial_{\mu} X - i g_{BL} n^{X}_{BL} Z_{BL} X$ where 
$n^{X}_{BL}$ is the $U(1)_{B-L}$ 
charge of the field $X$, as shown in Tables \ref{tab1} and \ref{tab2},
$g_{BL}$ is the $U(1)_{B-L}$ gauge coupling, and $Z_{BL}$ is the $U(1)_{B-L}$ 
gauge field. Finally, $\mathcal{V}\left(\phi_{h},\phi_{1}, \phi_{2} \right)$ is the potential consists of three scalars expressed as,
\begin{eqnarray}
\mathcal{V}(\phi_{h},\phi_{1},\phi_{2} ) & = &
- \mu^{2}_{h} \left( \phi^{\dagger}_{h} \phi_{h} \right) +
\lambda_{h} \left( \phi^{\dagger}_{h} \phi_{h} \right)^{2}
- \mu^{2}_{1} \left( \phi^{\dagger}_{1} \phi_{1} \right) +
\lambda_{1} \left( \phi^{\dagger}_{1} \phi_{1} \right)^{2}
- \mu^{2}_{2} \left( \phi^{\dagger}_{2} \phi_{2} \right) 
\nonumber \\
&+&
\lambda_{2} \left( \phi^{\dagger}_{2} \phi_{2} \right)^{2} + \lambda_{h1} \left( \phi^{\dagger}_{h} \phi_{h} \right)
\left( \phi^{\dagger}_{1} \phi_{1} \right)
+ \lambda_{h2} \left( \phi^{\dagger}_{h} \phi_{h} \right)
\left( \phi^{\dagger}_{2} \phi_{2} \right) \nonumber \\
&+& \lambda_{12} \left( \phi^{\dagger}_{1} \phi_{1} \right)
\left( \phi^{\dagger}_{2} \phi_{2} \right)
+ \mu \left( \phi_{2} \phi^{\dagger\,2}_{1} 
+ \phi^{\dagger}_{2} \phi^{2}_{1} \right)
\end{eqnarray}
\begin{itemize}
    
    \item {\bf Symmetry breaking:}

We first discuss the spontaneous symmetry breaking of the SM and $U(1)_{B-L}$
gauge groups when the scalars take the spontaneous vevs as follows,
\begin{eqnarray}
\phi_{h} =
\begin{pmatrix}
G^{+} \\
\frac{v+h + i G^{0}}{\sqrt{2}} 
\end{pmatrix},\quad
\phi_{1} = \frac{v_{1} + H_{1} + i A_{1}  }{\sqrt{2}},\quad
\phi_{2} = \frac{v_{2} + H_{2} + i A_{2}  }{\sqrt{2}}.
\end{eqnarray} 
As discussed in detail in Ref. \cite{Khan:2025yko}, we will have three physical CP-even 
Higgs state, one CP-odd scalar and the other {\it d.o.f} of the scalars become the longitudinal components of the SM and additional gauge bosons. The mass matrix for the CP-even states can be expressed after taking the second derivative
of the potential by obeying the tadpole conditions as follows on the basis 
$(h\,\,H_{1}\,\,H_{2})$,
\begin{eqnarray}
M^2_{scalar} = 
\begin{pmatrix}
2 \lambda_{h} v^2_{h} & \lambda_{h1} v_{h} v_{1} & \lambda_{h2} v_{h} v_{2} \\
\lambda_{h1} v_{h} v_{1} & 2 \lambda_{1} v^2_{1} 
& v_{1} \left( \sqrt{2} \mu + \lambda_{12} v_{2} \right) \\
\lambda_{h2} v_{h} v_{2} & v_{1} \left( \sqrt{2} \mu + \lambda_{12} v_{2} \right) & \left( -\frac{\mu v^2_{1}}{\sqrt{2} v_{2}} + 2 \lambda_{2} v^2_{2} \right)
\end{pmatrix}.
\end{eqnarray}
Therefore, the mass eigenstate, $(h_{1}\,\,h_{2}\,\,h_{3})$, and flavour 
eigenstate, $(h\,\,H_{2}\,\,H_{3})$, for the CP-even scalars can be related through the unitary 
mixing matrix, $U_{ij}$ ($i,j=1,2,3$), in the following way,
\begin{eqnarray}
\begin{pmatrix}
h\\
H_{1}\\
H_{2}
\end{pmatrix}
=   \smash{\underbrace{\begin{pmatrix}
c_{12} c_{13} & s_{12} c_{13} & s_{13} \\
-s_{12} c_{23} - c_{12} s_{23} s_{13} & c_{13} c_{23} - s_{12} s_{23} s_{13}
& s_{23} c_{13} \\
s_{12} s_{23} - c_{12} c_{23} s_{13} & -c_{12} s_{23} - s_{12} c_{23} s_{13}
& c_{23} c_{13}   
     \end{pmatrix}}_{U_{ij}}} 
\begin{pmatrix}
h_{1}\\
h_{2}\\
h_{3}
\end{pmatrix}.\\ \nonumber
\label{U-PMNS}
\end{eqnarray} 
In the same way, CP-odd scalars 
can be expressed in $2\times 2$ mass matrix in the basis 
$(A_{1}\,\,A_{2})$, as follows,
\begin{eqnarray}
M^2_{CP-odd} =
\begin{pmatrix}
 - 2 \sqrt{2} \mu v_{2} & \sqrt{2} \mu v_{1} \\
 \sqrt{2} \mu v_{1} & - \frac{\mu v_{1}}{\sqrt{2} v_{2}}
\end{pmatrix}.
\end{eqnarray}
One of the eigenvalues of the above mass matrix is zero, and the other one is 
non-zero corresponds to physical CP-odd scalar $A$ with masses as follows,
\begin{eqnarray}
M_{G_{BL}} = 0,\quad M^2_{A} = -2 \sqrt{2} \mu v_{2} \left( 1 + \frac{v^2_{1}}{4 v^2_{2}} \right)\,.
\end{eqnarray}
Finally, the mass eigenstates and flavour eigenstates for the CP-odd
scalars can be related as,
\begin{eqnarray}
\begin{pmatrix}
G_{BL} \\
A
\end{pmatrix}
=
\begin{pmatrix}
 \cos\beta & -\sin\beta \\
 \sin\beta & \cos\beta
\end{pmatrix}
\begin{pmatrix}
A_{1} \\
A_{2}
\end{pmatrix}
\end{eqnarray}
where $\tan\beta = \frac{2 v_{2}}{v_{1}}$.
In the above mass basis, $G_{BL}$ is the Goldstone boson associated with
the longitudinal component of $U(1)_{B-L}$ gauge boson $Z_{BL}$ having mass,
\begin{eqnarray}
M^2_{Z_{BL}} &=& g^2_{BL} v^2_{1} + 4 g^2_{BL} v^2_{2}\,\nonumber \\
 &=& g^2_{BL} v^2_{1} \left(1 + \tan^{2}\beta \right).
\end{eqnarray}

\item{\bf Yukawa terms for BSM fermions:}

The Yukawa terms associated with the BSM fermions $\psi_{1,2}$ 
and the Higgs bosons, $h_{1,2,3}$, after the symmetry breaking take the following form,
\begin{eqnarray}
\mathcal{L}^{Yuk}_{\psi} &=&
\sum_{i=1,2,3} \alpha_{11i} \bar \psi_{1L} \psi_{1R} h_{i} +
\sum_{i=1,2,3} \alpha_{12i} \bar \psi_{1L} \psi_{2R} h_{i} +  
\sum_{i=1,2,3} \alpha_{21i} \bar \psi_{2L} \psi_{1R} h_{i} 
\nonumber \\ &+&
\sum_{i=1,2,3} \alpha_{22i} \bar \psi_{2L} \psi_{2R} h_{i} 
+ \mathrm{\it i}\, \alpha_{11A}\, \bar \psi_{1L} \psi_{1R} A
+ \mathrm{\it i}\, \alpha_{12A}\, \bar \psi_{1L} \psi_{2R} A 
+ \mathrm{\it i} \, \alpha_{21A}\, \bar \psi_{2L} \psi_{1R} A
\nonumber \\
&+& \mathrm{\it i}\, \alpha_{22A}\, \bar \psi_{2L} \psi_{2R} A + {\it h.c.}\,.
\end{eqnarray}
where the vertex factor $\alpha_{ijk}$ for the Yukawa interactions take the following form,
\begin{eqnarray}
\alpha_{11i} &=& \frac{M_{1}}{\sqrt{2} v_{1} v_{2}} 
\left[ U_{3i} v_{1} + U_{2i} v_{2} + \left( U_{3i} v_{1} - U_{2i} v_{2} \right) \cos 2 \theta_{L} \right]\,, \nonumber \\
\alpha_{12i} &=& \frac{\sqrt{2} M_{2}}{v_{1} v_{2}} 
\left[  \left( U_{3i} v_{1} - U_{2i} v_{2} \right) \cos \theta_{L}
\sin\theta_{L} \right]\,, \nonumber \\
\alpha_{21i} &=& \frac{\sqrt{2} M_{1} }{ v_{1} v_{2} } 
\left[  \left( U_{3i} v_{1} - U_{2i} v_{2} \right) \cos\theta_{L}
\sin\theta_{L} \right]\,, \nonumber \\
\alpha_{22i} &=& \frac{M_{2}}{\sqrt{2} v_{1} v_{2}} 
\left[ U_{3i} v_{1} + U_{2i} v_{2} + \left( -U_{3i} v_{1} + U_{2i} v_{2} \right) \cos 2 \theta_{L} \right].\,
\label{psipsihi}
\end{eqnarray}
Here, $U_{ij}$ is the Higgses mixing matrix defined in Eq. (\ref{U-PMNS}), 
and $U_{2A} = \sin\beta$, $U_{3A} = \cos\beta$ for $j=A$ and
$\theta_L$ is the mixing angle between the extra Dirac fermions, which can modify the decay width of the NLSP to DM. 

Finally, on the mass basis of BSM fermions, we can write down the interaction
between the BSM fermions and the gauge boson in the following way, 
\begin{eqnarray}
\mathcal{L}_{\psi Z_{BL}} &=& - \frac{g_{BL}}{3} 
\biggl[ \bar \psi_{1} \gamma^{\mu} \left( (3 \cos^{2}\theta_{L} +1)P_{L}
- 2 P_{R} \right) \psi_{1}
+ \bar \psi_{2} \gamma^{\mu} \left( (3 \sin^{2}\theta_{L} +1)P_{L}
- 2 P_{R} \right) \psi_{2} \nonumber \\
&+& \bar \psi_{1} \gamma^{\mu} (2 \sin^{2}\theta_{L}) P_{L} \psi_{2}
+ \bar \psi_{2} \gamma^{\mu} (2 \sin^{2}\theta_{L}) P_{L} \psi_{1}
 \biggr] Z_{BL\,\mu}\,.
\end{eqnarray}
where $P_{L/R} = \frac{1\mp \gamma_5}{2}$ are the projection operators.

\end{itemize}
\section{Constraints}
\label{constraints}

In this section, we briefly describe all the relevant constraints 
which are relevant for our study.  
In particular, we have considered theoretical bounds, namely perturbativity and vacuum stability bounds. 
The bounds on DM coming from its relic density, direct and indirect detection 
experiments.
The collider bounds, mainly on the  Higgses mixing angles coming
from the Higgs signal strength and branching to 
invisible particles, 
which also ensures the safety of the present model from the oblique parameters bound.
Moreover, for long-lived particle NLSP $\psi_2$ decay, we have also discussed the 
BBN and CMB bounds. 

\subsection{Perturbativity and vacuum stability bounds}

The perturbativity conditions on the quartic and Yukawa couplings ensure the validity of the theory at leading order. The quartic couplings can be expressed in terms of the physical scalar masses and scalar mixing angles as follows,
\begin{eqnarray}
\lambda_{h} &=& \frac{U^2_{11} M^2_{h_{1}} + U^2_{12} M^2_{h_{2}} 
+ U^2_{13} M^2_{h_{3}} }{2 v^2_{h}}\,,\nonumber \\
\lambda_{1} &=& \frac{U^2_{21} M^2_{h_{1}} + U^2_{22} M^2_{h_{2}} 
+ U^2_{23} M^2_{h_{3}} }{2 v^2_{1}}\,, \nonumber \\
\lambda_{2} &=& \frac{U^2_{31} M^2_{h_{1}} + U^2_{32} M^2_{h_{2}} 
+ U^2_{33} M^2_{h_{3}} - M^2_{A} \cos^{2}\beta  }{2 v^2_{2}} \,,\nonumber\\
\lambda_{h1} &=& \frac{U_{11} U_{21} M^2_{h_{1}} 
+ U_{12} U_{22} M^2_{h_{2}} + U_{13} U_{23} M^2_{h_{3}} }{v_{h}v_{1} }
\,, \nonumber \\
\lambda_{h2} &=& \frac{U_{11} U_{31} M^2_{h_{1}} 
+ U_{12} U_{32} M^2_{h_{2}} + U_{13} U_{33} M^2_{h_{3}} }{v_{h}v_{2} }
\,, \nonumber \\
\lambda_{12} &=& \frac{U_{21} U_{31} M^2_{h_{1}} 
+ U_{22} U_{32} M^2_{h_{2}} + U_{23} U_{33} M^2_{h_{3}} + M^2_{A}
\sin\beta\, \cos\beta }{v_{1}v_{2} }
\,. \nonumber \\
\label{quartic-coupling-expression}
\end{eqnarray}
In our analysis, we have required that the quartic couplings remain within the range 
0 to $4\pi$. Some of the mixed quartic couplings can be negative, and their minimum values are determined by the behaviour of the potential at large field values. By demanding that the potential remains positive at large field values as well, we obtain the following bounds,
\begin{eqnarray}
&&\lambda_{h} > 0\,,\,\, \lambda_{1} > 0\,,\,\, \lambda_{2} > 0\,,\,\,
 \lambda_{12} + 2 \sqrt{\lambda_{1} \lambda_{2} } > 0\,,
 - 2 \sqrt{\lambda_{h} \lambda_{1}} \leq \lambda_{h1} \leq 2 \sqrt{\lambda_{h} \lambda_{1}}\,,\nonumber \\
&& - 2 \sqrt{\lambda_{h} \lambda_{2}} \leq \lambda_{h2} \leq 2 \sqrt{\lambda_{h} \lambda_{2}}\,,
2 \lambda_{h} \lambda_{12} - \lambda_{h1} \lambda_{h2} 
+ \sqrt{\left( 4 \lambda_{h} \lambda_{1} -\lambda^2_{h1}  \right)
\left( 4 \lambda_{h} \lambda_{2} -\lambda^2_{h2}  \right)} \geq 0\,.
\end{eqnarray}
It is to be noted that we have always taken the quartic couplings positive, so the above bounds are trivially satisfied.

\subsection{DM relic density}

In the present work, we have produced the DM by the CDFO
and the NLSP decays to DM, similar to the superWIMP mechanism. During our scanning,
to get more data points in a reasonable time, we have demanded the DM relic
density from $(1-100)\%$, which translates to the DM relic density in the following range,
\begin{eqnarray}
10^{-3} \leq \Omega^{tot}_{\psi_{1}} h^{2} \left(= \underbrace{\Omega_{\psi_{1}}h^{2}}_{Conversion}
 + \overbrace{\frac{M_{\psi_1}}{M_{\psi_2}} \Omega_{\psi_{2}} h^{2}}^{superWIMP} \right)  \leq 0.1284\,.
\end{eqnarray}
It is to be noted that by a slight change of the model parameters, mainly the NLSP interaction with the visible sector and the mass difference between NLSP and DM, we can easily get 100\% of the DM, as presented in
Planck 2018 data \cite{Planck:2018vyg}, without affecting the phenomenology and
other constraints.

\subsection{DM direct detection bounds}
\begin{figure}[h!]
\centering
\includegraphics[angle=0,height=4.5cm,width=14.5cm]{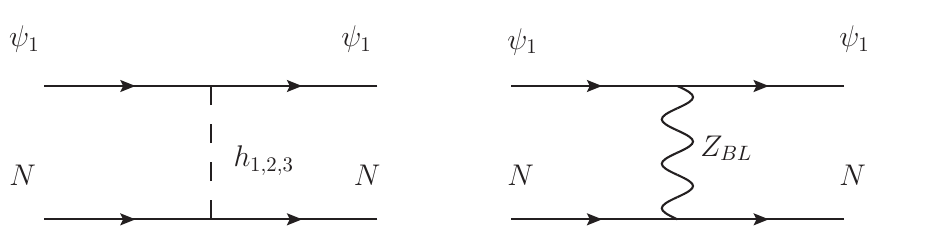}
\caption{DM direct detection diagrams relevant in our study.} 
\label{DM-DD}
\end{figure}
The DM direct detection cross section for $\psi_{1} N \rightarrow \psi_{1} N$
as shown in Fig. \ref{DM-DD} takes the following form,
\begin{eqnarray}
\sigma_{\psi_{1}} = \frac{\mu^{2}}{\pi} \left[ \frac{f_{N} M_{N}}{v} 
\sum_{i = 1,2,3} \frac{U_{1i} \alpha_{11i}}{M^2_{h_{i}}} 
+ \frac{f_{Z_{BL}} g^{2}_{BL} \left( 3 \cos^{2}\theta_{L} -1 \right) }{18 M^2_{Z_{BL}}} \right]^{2}
\label{DD-expression}
\end{eqnarray} 
where $\mu = \frac{M_{\psi_{1}} M_{N}}{M_{\psi_{1}} + M_{N}}$ and 
$f_{N} \sim 0.3$ \cite{Cline:2013gha}, $f_{\zmt} = 3$ \cite{Belanger:2008sj}. 
In our study, we have considered DM density in the ($1-100$)\% range and we can
easily tune the parameters to get 100\% DM and it will not affect the DM
direct detection. This is because in our DM production by CDFO, 
the DM density will be controlled by the 
NLSP interaction with the SM particles, therefore, without changing the DM direct detection cross section, we can achieve 100\% DM density. 
We have shown bounds coming from recent LUX-ZEPLIN 
direct detection data \cite{LZCollaboration:2024lux} as well 
as the future DARWIN projection \cite{DARWIN:2016hyl}.

\subsection{DM indirect detection bounds}

\begin{figure}[h!]
\centering
\includegraphics[angle=0,height=4.5cm,width=14.5cm]{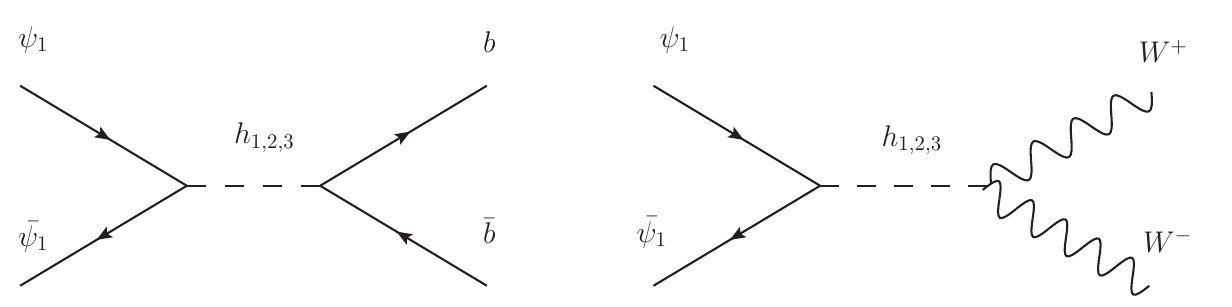}
\caption{DM indirect detection diagrams relevant in our study.} 
\label{indirect-detection}
\end{figure}
The DM in the present work can also be detected by the indirect detection 
experiments through the diagrams shown in Fig. \ref{indirect-detection}.
There are many indirect detection experiments of DM which are looking for DM signal 
indirect way, namely Fermi-LAT \cite{MAGIC:2016xys} and many future experiments in this direction, such as the Cherenkov Telescope Array \cite{CTA:2020qlo}. The indirect detection cross section in the non-relativistic limit 
can be expressed as, 
\begin{eqnarray}
\left( \sigma v \right)_{kk} &\simeq& \frac{n_{c} v^{2}_{rel} M^2_{b} M^2_{\psi_{1}} 
\left( 1 - \frac{M^2_{b}}{M^2_{\psi_{1}}} \right)^{3/2}}{8 \pi v^{2}} \sum_{i,j = 1,2,3}A_{i} A^{*}_{j}\,,\,\,{\rm for}\,\,k=b\,,\nonumber \\  
 &\simeq& \frac{v^2_{rel} M^4_{W} \sqrt{1 - \frac{M^2_{W}}{M^2_{\psi_{1}}}} }{16 \pi v^{2}} \left(3 - \frac{4 M^2_{\psi_{1}}}{M^2_{W}} 
 + \frac{4 M^4_{\psi_{1}}}{M^4_{W}} \right) \sum_{i,j = 1,2,3}A_{i} A^{*}_{j}  
\,,\,\,{\rm for}\,\,k=W^{\pm}\,.
\label{ID-analytical}
\end{eqnarray}  
where $v_{rel} \sim 10^{-3}$ is the DM velocity in the Galactic center and
$n_{c} = 3$ is a colour charge of the final state particles.
The amplitude $A_i$ takes the form,
\begin{eqnarray}
A_{i} = \frac{\alpha_{11i} U_{1i}}{\left( 4 M^2_{\psi_{1}} - M^2_{h_{i}} \right) 
+ i \Gamma_{h_{i}} M_{h_{i}} }\,.
\end{eqnarray}
Our WIMP DM is fermionic in nature, and direct interaction with the visible sector has 
been chosen to be small
contrary to the usual freeze-out scenario. Therefore, our indirect detection cross
section is much below than the present sensitivity as well as the future 
proposed indirect detection experiment Cherenkov Telescope Array \cite{CTA:2020qlo}.

\subsection{Collider bounds}
The collider bounds in the present work mainly come from the Higgs 
observables measurements. In this context, Higgs signal strength and Higgs invisible branching are the important ones. As shown in 
Refs. \cite{Heo:2024cif, Khan:2025yko},
the Higgs signal strength puts a bound on the mixing angle as $\sin {\theta_{12}} \leq 0.23$, assuming $\theta_{13}$ is small. Moreover, we have also taken 
into account the bound from the Higgs invisible decay mode, 
{\it i.e.} $Br_{inv} < 0.16$
at $95\%$ confidence level \citep{CMS:2018yfx, CMS:2021far, CMS:2020ulv},
when the DM mass is smaller than half of the SM Higgs mass.

Moreover, the collider bounds on the additional gauge boson $Z_{BL}$, primarily 
from LEP \cite{Carena:2004xs} and LHC \cite{Cacciapaglia:2006pk, CMS:2012umo, 
ATLAS:2014pcp}, are not relevant for our analysis, as we consider the regime where the additional gauge coupling is small compared to the 
weak gauge coupling.

\subsection{BBN bounds}
In the present work, we have studied the single-component DM produced by the 
conversion-driven freeze-out, and its density depends on the density 
of the NLSP abundance. Additionally, at a later time, NLSP can decay to the DM candidate
along with some visible particle around the BBN time. This may pose a 
problem to the successful prediction of the BBN. Therefore, we have constrained our
parameter space by following Ref. \cite{Kawasaki:2017bqm} and indicate the parameter
values which can lead to this kind of situation.
Moreover, the additional gauge boson decays much before the BBN time for the choice 
of our range of parameters to realise the CDFO DM.
 
Finally, as shown in detail in Ref. \cite{Khan:2025yko}, the present model
is safe from the bounds coming from the oblique parameters
$ S, T, U$ \cite{CDF:2013dpa, ATLAS:2017rzl, LHCb:2021bjt, CDF:2013bqv}
 mainly because of the tight constraints on the Higgs mixing angles 
coming from the Higgs signal strength measurements.

\section{Results}
\label{result}

\begin{figure}[h!]
\centering
\includegraphics[angle=0,height=7.5cm,width=12.5cm]{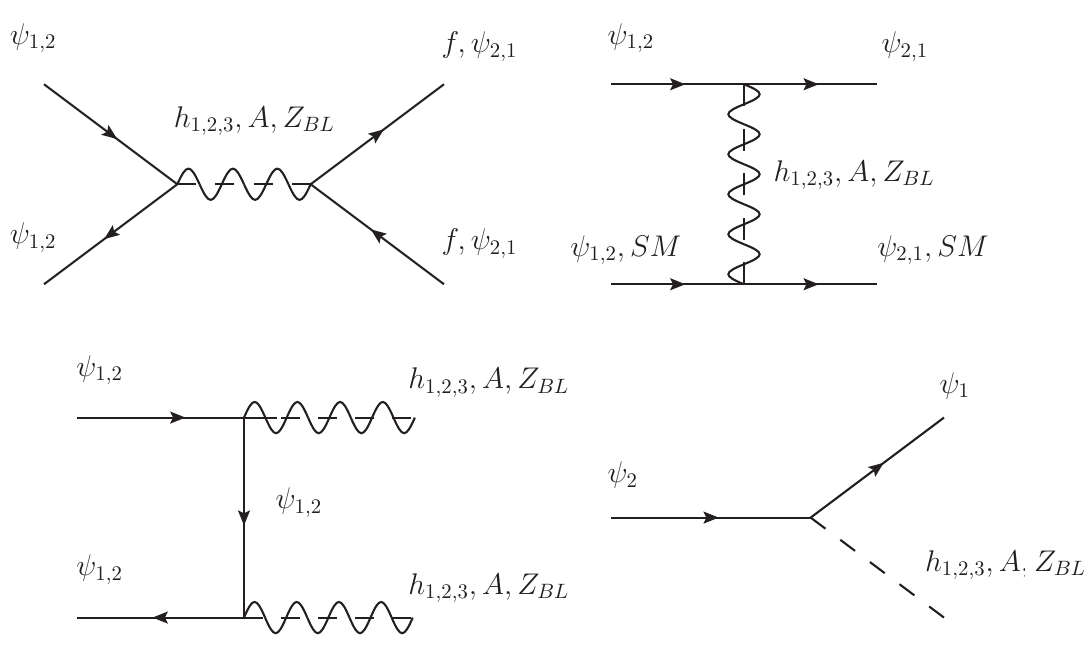}
\caption{Diagrams relevant in setting the DM and NLSP abundances.} 
\label{DM-annihilation-diagram}
\end{figure}

In the present work, we have considered $\psi_2$ as the NLSP and $\psi_1$ as the DM candidate.
The coupled Boltzmann equations for
$\psi_1$ and $\psi_2$ candidates can be expressed as,
\begin{eqnarray}
&\frac{d Y_{\psi_1}}{dz}& = -\frac{s}{H z} \biggl[ 
\langle\sigma v\rangle_{\psi_{1} \psi_1 \rightarrow f \bar{f}}
\biggl( Y^2_{\psi_1} - (Y^{eq}_{\psi_1})^{2} \biggr)
+ \langle \sigma v \rangle_{\psi_{1} \psi_2 \rightarrow f \bar{f}}
\biggl( Y_{\psi_1} Y_{\psi_2} - Y^{eq}_{\psi_1} Y^{eq}_{\psi_2}  \biggr)
\nonumber \\
&-& \frac{\Gamma_{\psi_2 \rightarrow \psi_{1}}}{s(T)} 
\biggl( Y_{\psi_{2}} - Y_{\psi_1} \frac{Y^{eq}_{\psi_2}}{Y^{eq}_{\psi_1}} \biggr) - \langle \sigma v  \rangle_{\psi_{2} \psi_{2} \rightarrow 
\psi_{1} \psi_{1}} \biggl( Y^2_{\psi_2} - \frac{(Y^{eq}_{\psi_2})^{2}}
{(Y^{eq}_{\psi_1})^{2}} Y^2_{\psi_1} \biggr)  
- \langle \sigma v \rangle_{\psi_{1} \psi_{2} \rightarrow 
\psi_{1} \psi_{1}} \biggl( Y_{\psi_1} Y_{\psi_2} - 
\frac{Y^{eq}_{\psi_2}}{Y^{eq}_{\psi_1}} Y^2_{\psi_1} \biggr)
 \nonumber \\
&-&
\langle \sigma v \rangle_{\psi_{2} \psi_{2} \rightarrow 
\psi_{1} \psi_{2}} \biggl( Y^2_{\psi_2} - 
\frac{Y^{eq}_{\psi_2}}{Y^{eq}_{\psi_1}} Y_{\psi_1} Y_{\psi_2} \biggr)
- \frac{\widetilde{\Gamma}_{\psi_2 \rightarrow \psi_1 X  } }{s(T)}
\biggl( Y_{\psi_2} - \frac{Y^{eq}_{\psi_2}}{Y^{eq}_{\psi_1} } Y_{\psi_1} 
\biggr) 
\biggr], \nonumber \\
&\frac{d Y_{\psi_2}}{dz}& = -\frac{s}{H z} \biggl[ 
\langle\sigma v\rangle_{\psi_{2} \psi_2 \rightarrow f \bar{f}}
\biggl( Y^2_{\psi_2} - (Y^{eq}_{\psi_2})^{2} \biggr)
+ \langle \sigma v \rangle_{\psi_{1} \psi_2 \rightarrow f \bar{f}}
\biggl( Y_{\psi_1} Y_{\psi_2} - Y^{eq}_{\psi_1} Y^{eq}_{\psi_2}  \biggr)
\nonumber \\
&+& \frac{\Gamma_{\psi_2 \rightarrow \psi_{1} }}{s(T)} 
\biggl( Y_{\psi_{2}} - Y_{\psi_1} \frac{Y^{eq}_{\psi_2}}{Y^{eq}_{\psi_1}} \biggr) + \langle \sigma v  \rangle_{\psi_{2} \psi_{2} \rightarrow 
\psi_{1} \psi_{1}} \biggl( Y^2_{\psi_2} - \frac{(Y^{eq}_{\psi_2})^{2}}
{(Y^{eq}_{\psi_1})^{2}} Y^2_{\psi_1} \biggr)  
+ \langle \sigma v \rangle_{\psi_{1} \psi_{2} \rightarrow 
\psi_{1} \psi_{1}} \biggl( Y_{\psi_1} Y_{\psi_2} - 
\frac{Y^{eq}_{\psi_2}}{Y^{eq}_{\psi_1}} Y^2_{\psi_1} \biggr)
 \nonumber \\
&+&
\langle \sigma v \rangle_{\psi_{2} \psi_{2} \rightarrow 
\psi_{1} \psi_{2}} \biggl( Y^2_{\psi_2} - 
\frac{Y^{eq}_{\psi_2}}{Y^{eq}_{\psi_1}} Y_{\psi_1} Y_{\psi_2} \biggr)
+ \frac{\widetilde{\Gamma}_{\psi_2 \rightarrow \psi_1 X  } }{s(T)}
\biggl( Y_{\psi_2} - \frac{Y^{eq}_{\psi_2}}{Y^{eq}_{\psi_1} } Y_{\psi_1} 
\biggr) 
\biggr].
\label{BE}
\end{eqnarray}
In the above Boltzmann equation, we have 
defined $z = \frac{M_{\psi_1}}{T}$ and the velocity times cross section 
$\langle\sigma{}v\rangle_{AB\rightarrow CD}$ corresponds for the process $AB\rightarrow CD$
as shown in Fig. \ref{DM-annihilation-diagram}. 
The interaction rate $ \Gamma_{ \psi_{2} \rightarrow \psi_{1} } = \sum_{f = SM\, states}
\langle \sigma v \rangle_{\psi_2 f \rightarrow \psi_1 f} n^{eq}_{f}$ implies
the conversion rate between $\psi_2 \leftrightarrow \psi_1$ as long as the process is in thermal equilibrium.
The thermal average decay width $\widetilde{\Gamma}_{\psi_{2}\rightarrow \psi_{1} X } =
 \frac{K_{2}\biggl(\frac{M_{\psi_2}}{M_{\psi_1}} z \biggr)}
 {K_{1}\biggl(\frac{M_{\psi_2}}{M_{\psi_1}} z \biggr)} \Gamma_{\psi_{2}\rightarrow \psi_{1} X }$ where $K_{1,2}(z)$ are 
 the modified Bessel function of the second kind and 
  $X = A, Z_{BL}, h_{1,2,3}$ can be on shell or off shell depending 
  on the masses of $\psi_{1}$ and $\psi_2$. The equilibrium co-moving 
  number density of $\psi_{1}$ and $\psi_2$ can be expressed as,
  \begin{eqnarray}
  Y^{eq}_{\psi_1} = \frac{45 z^{2}}{2 \pi^{2} g_{s}(M_{\psi_1}/z)} K_{2}(z),\quad Y^{eq}_{\psi_2} = \frac{45 z^{2}}{2 \pi^{2} g_{s}(M_{\psi_1}/z)} 
  K_{2}\biggl(\frac{M_{\psi_2}}{M_{\psi_1}}z\biggr)
\end{eqnarray}   
In solving the Boltzmann equations in Eq. \ref{BE}, we have 
used publicly available package micrOMEGAs (v6.1.5) \cite{Alguero:2023zol}.
In particular, we have used the routine "darkOmegaNTR($T_R$, Y, fast, Beps, \&err)" for determining the evolution of our DM and set the global variable $Tend = 10^{-3}$ GeV
which basically implies that we have the information of the comoving number
density $Y[i]$ ($i$ is the DM ordering for multi-component DM scenario) 
in between the initial temperature $T_R$ and final temperature $Tend$.

\subsection{Line plots}
\label{line-plots-dm}

\begin{figure}[h!]
\centering
\includegraphics[angle=0,height=6.5cm,width=7.5cm]{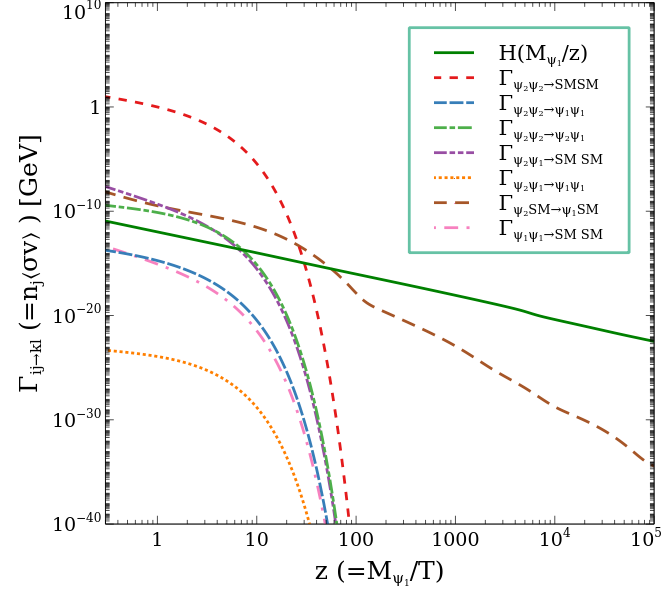}
\includegraphics[angle=0,height=6.5cm,width=7.5cm]{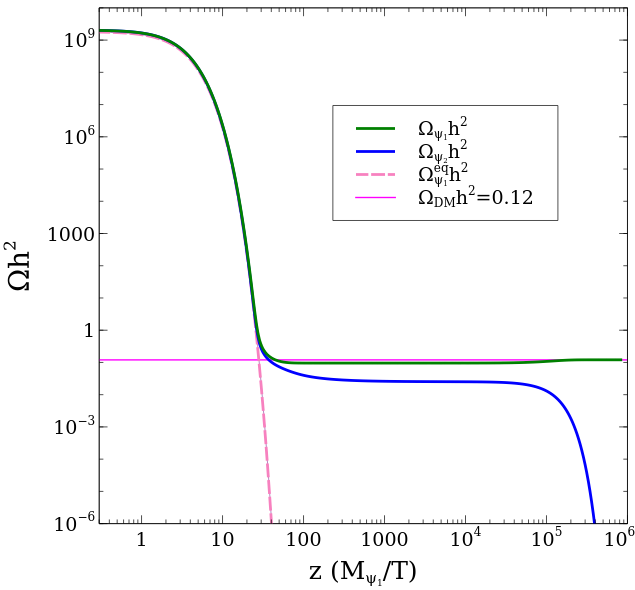}\\
\includegraphics[angle=0,height=6.5cm,width=7.5cm]{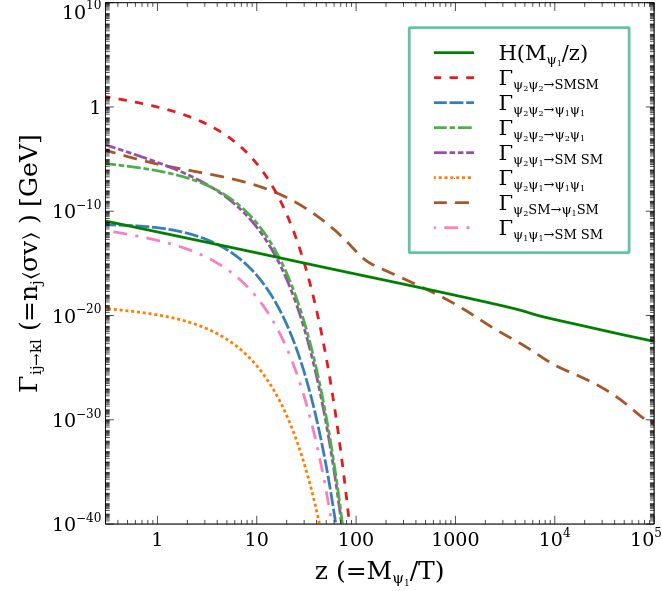}
\includegraphics[angle=0,height=6.5cm,width=7.5cm]{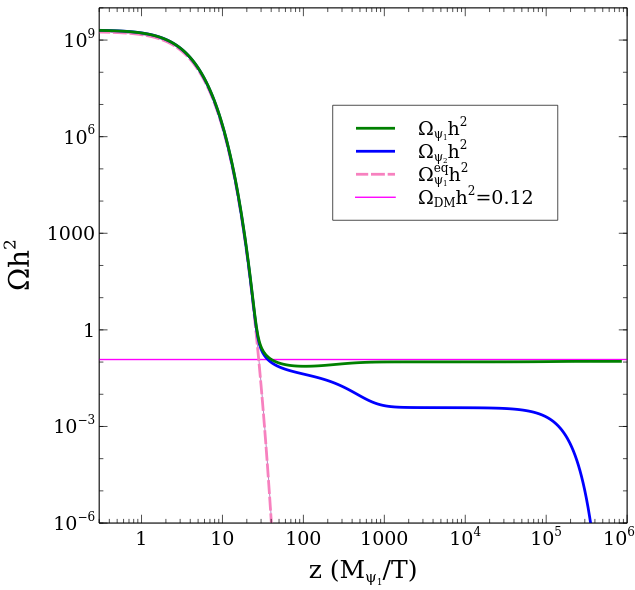}
\caption{In the LP and RP, we have shown the variation of interaction
rates of $\psi_{1,2}$ and DM relic density with z. 
In generating the upper two plots, we have kept the model parameter values fixed at the following values
$\alpha_{111} = -6.13\times10^{-9}$, $\alpha_{112} =
 9.94\times10^{-8}$, $\alpha_{113} = 8.07\times10^{-8}$, $\alpha_{12} = 
 \alpha_{121}
 = - \alpha_{122} = - \alpha_{123} = \alpha_{211} = -\alpha_{212}
 = - \alpha_{213} = - \alpha_{12A} = - \alpha_{21A} = 4.5\times10^{-6}$, $\alpha_{221} = 
 -1.05\times10^{-5}$, $\alpha_{222} = 1.036$, $\alpha_{223} = 
 1.438\times10^{-4}$, $\alpha_{11A} = 7.939\times10^{-8}$, 
 $\alpha_{22A} = 1.0359$, $M_{\psi_1} = 860.098$ GeV, 
 $M_{\psi_2} = 865.004$ GeV, $M_{h_2} = 633.31$ GeV, 
 $M_{h_3} = 776.79$ GeV, $M_{A_2} = 87.45$ GeV, $g_{D} = 
 6.2 \times 10^{-8}$, $M_{Z_D} = 1.93$ GeV, $\tan\beta = 2.47$ and
 the Higgs mixing matrix is diagonal {\it i.e.} $U_{ij} \simeq 
 \mathbb{I}$ 
 ($i,j = 1,2,3$). In the lower two plots, we have taken $|\alpha_{12i}| =
 |\alpha_{21i}| = 4.5 \times 10^{-4}$ ($i=1,2,3$). } 
\label{line-plot-1}
\end{figure}

In Fig. \ref{line-plot-1}, we have shown the variation of the DM ($\psi_1$) 
and NLSP ($\psi_2$) interaction rates with $z$ in the 
left-hand side plots, whereas in the right-hand side plots, we have shown the variation of DM and NLSP relic densities. The parameter values have been provided in the caption of the figure. In particular, the upper panel plots correspond to $|\alpha_{12i}| = |\alpha_{21i}| = 4.5 \times 10^{-6}$ ($i=1,2,3,A$), and the lower panel plots correspond to $|\alpha_{12i}| = |\alpha_{21i}| = 4.5 \times 10^{-4}$. Additionally, the exact form of the Higgs mixing matrix $U_{ij}$ has been shown in Eq. \ref{higgs-mixing-matrix} in the Appendix.
In determining the thermal average cross section, $\langle\sigma v\rangle_{abcd}$
where $a,b,c,d = 0,1,2$ based on SM and ordering of the DM components, 
we have used micrOMEGAs in-built
routine ``vSigmaNCh(T, "abcd", fast, Beps, $\langle\sigma v\rangle_{abcd}$)''
 where $T$ is the temperature where it is measured, and other things are 
discussed in the micrOMEGAs manual, which are related to numerical precision. 
In determining the DM/NLSP equilibrium number density, we have used 
micrOMEGAs in built function $n_{\alpha}(T) = \frac{YdmNEq(T,\alpha)}{s(T)}$
where $\alpha = \psi_{1}, \psi_2$ and the entropy density is define as 
$s(T) = \frac{2 \pi^{2}}{45} g_{*}(T) T^{3}$ where $g_{*}(T)$ is the total 
relativistic {\it d.o.f} of the Universe. Moreover, in determining the number density
of the SM particle, we have used the relativistic expression of the number
density by taking into account all the SM {\it d.o.f} \footnote{For high 
temperature with SM particle content,
the total number of relativistic d.o.f is $g_{*} = 106.75$.} which reduced with
the temperature {\it i.e.} $n_{SM} = \frac{1.2}{\pi^2} g_{*}(T) T^3$.  
In the upper plots of Fig. \ref{line-plot-1}, we can see that $\psi_2 \psi_2 \rightarrow
SM~SM$, $\psi_2 \psi_1 \rightarrow SM~SM$,
$\psi_2~SM \rightarrow \psi_1~SM$ processes keep $\psi_2$ and $\psi_1$
particles in thermal equilibrium. 
The annihilation processes $\psi_1 \psi_1 \rightarrow SM~SM$ and $\psi_i \psi_j \rightarrow \psi_i \psi_j$
 are below the Hubble rate. The 
$\psi_{1} \psi_1 \leftrightarrow SM~SM$ is below the Hubble rate in both the plots.
Even without the effective direct annihilation to
two SM particles, our DM can be in thermal equilibrium and achieve 
the freeze-out when $\psi_2~SM \leftrightarrow \psi_1~SM$ goes out of 
equilibrium. It is to be noted that $\psi_2$ NLSP is in chemical 
equilibrium with the SM bath and goes out of equilibrium at late time, as
shown in the plots.  
The number density
of $\psi_{1}$ DM will depend on when the process $\psi_2 \psi_2 
\leftrightarrow SM~SM$ goes out of equilibrium. The most important thing of the present set-up is that we do not need the processes
$\psi_{1} \psi_{1,2} \leftrightarrow SM~SM$ to be in thermal equilibrium, in contrast to the usual freeze-out mechanism of DM.
In the RP, we have shown the relic density variation of $\psi_{1}$ and $\psi_2$. 
We can see that $\psi_1$ goes out of equilibrium when $\psi_2$ goes out of equilibrium from the SM bath, and then they interchange their abundances through $\psi_2~ SM \leftrightarrow \psi_1~SM$
until the process goes out of equilibrium completely. 
Depending on the mass difference, $\psi_1$ and $\psi_2$ freezes out to particular values, and later on, $\psi_2$ decays and 
converts its density to $\psi_1$ DM through the superWIMP mechanism.
In the lower plot, we have shown the results for higher values of 
$|\alpha_{12i}|$ and $|\alpha_{21i}|$, which lead to higher interaction rates 
involving $\psi_1$ and $\psi_2$ annihilation and co-annihilation compared to the previous case, as seen in the figure. Since the coupling values are higher, we observe that $\psi_2~ SM \rightarrow \psi_1~SM$ remains above the Hubble rate 
for a longer time, like $z\simeq 450$. We can see a increment in the 
process $\psi_1 \psi_1 \leftrightarrow SM~SM$ which mainly happen for the 
process $\psi_1 \psi_1 \xrightarrow{\psi_2} h_{i} h_{i}$ which depends on
$|\alpha_{12i}|$ ($i=1,2,3$).
In the RP, we again show the relic density variation of $\psi_1$ and $\psi_2$. After the freeze-out of $\psi_{2}$ which is unchanged, $\psi_{1,2}$ keeps interacting each other
and freezes out later on, and eventually, NLSP decays into DM. 
In both plots, $\psi_2$ decays mainly via $\psi_{1} Z_{BL}$, 
and the associated parameters are kept fixed in the upper and lower panel plots, 
so in both cases $\psi_2$ decays at the same time. 

\begin{figure}[h!]
\centering
\includegraphics[angle=0,height=7cm,width=7.5cm]{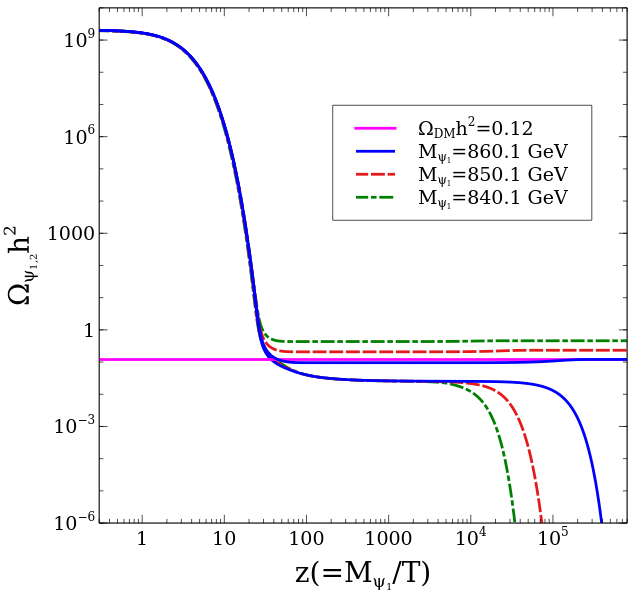}
\includegraphics[angle=0,height=7cm,width=7.5cm]{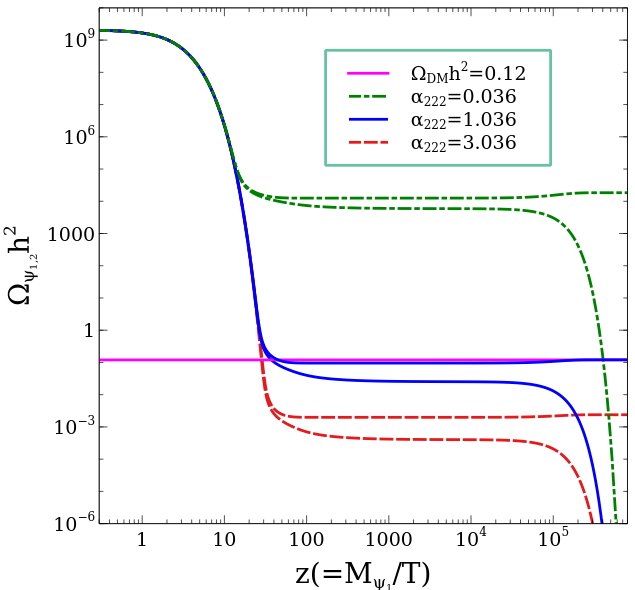}
\caption{The LP shows the variation of DM relic density with $z$ for three values of the DM mass $M_{\psi_1}$, whereas in the RP, the variation is shown for three values of $\alpha_{222}$, which controls the $\psi_2$ interaction with the visible sector. The other parameter values have been kept fixed at the values shown in the caption of Fig. \ref{line-plot-1}.} 
\label{line-plot-2}
\end{figure}

In the LP of Fig. \ref{line-plot-2}, we have shown the change in the DM relic density for three values of the DM mass $M_{\psi_1}$. From the plot, it can be seen that as the mass $M_{\psi_1}$ decreases compared to the fixed NLSP mass
$M_{\psi_2} = 865.0$ GeV, we observe an earlier freeze-out of $\psi_1$, resulting in an increased of $\psi_1$ density. This occurs because the 
interaction rate $\psi_1~SM \rightarrow \psi_2~SM$ is suppressed by the 
factor $e^{-(M_{\psi_2} - M_{\psi_1})/T}$, causing $\psi_1$ to freeze out earlier as the mass difference between $\psi_2$ and $\psi_1$ increases.
At the same time, due to the increased mass difference, the interaction rate 
of $\psi_2$ with the SM bath remains unaffected, which is mainly controlled by
$\psi_{2} \psi_{2} \xrightarrow{\psi_2} h_{2} h_{2}$, so $\psi_2$ freezes out at the same time with the same abundance for all three values. 
We also observe that $\psi_2$ decays occur earlier as the mass difference increases. 
This is because the dominant 
decay mode\footnote{In micrOMEGAs, if the two-body decay mode is open, then it does not take into account the three-body decay 
mode. In our study, the three-body decay is controlled by 
$\alpha_{12i}$ and the two-body decay by $g_{BL}$, so depending on their strength, the three-body decay can dominate over the two-body decay, and we have taken this 
into account during the MATHUSLA and BBN study in 
sections (\ref{mathusla-section}, \ref{bbn-section}).}, for the choice of parameter range, 
$\psi_2 \rightarrow \psi_1 Z_{BL}$ 
is proportional to $\left(M_{\psi_2} - M_{\psi_1}\right)^3$ in the limit $M_{\psi_{2,1}} \gg M_{Z_{BL}}$.
In the RP of Fig. \ref{line-plot-2}, we have shown the variation of NLSP 
and DM densities for three values of the coupling $\alpha_{222}$, which controls 
the interaction of $\psi_2$ with the SM bath. Depending 
on the value of $\alpha_{222}$, NLSP decouples 
earlier or later from the thermal bath with higher or lower
number density. 
The interaction of 
$\psi_2$ with $h_2$ is not suppressed by any Higgs mixing angle, 
so the parameter $\alpha_{222}$ directly controls the strength of 
$\psi_2$'s interactions through the process $\psi_{2} \psi_{2}
 \xrightarrow{\psi_2} h_{2} h_{2}$, as reflected in the figure.
Once NLSP $\psi_2$ decouples then $\psi_2~SM \leftrightarrow 
\psi_{1}~SM$ keeps taking place and decouples from the bath later on. 
As we can see from the figure, when $\alpha_{222}$ decreases, we observe 
earlier freeze-out of both $\psi_2$ and $\psi_1$, 
resulting in a larger DM abundance, which mainly happens because of
earlier decoupling of $\psi_2$ with higher abundance. 
The plot implies that by varying $\alpha_{222}$ which is not associated with
$\psi_1$, we can achieve the desired DM abundance while 
keeping the $\psi_1$ interaction with the SM at an arbitrary strength.

\begin{figure}[h!]
\centering
\includegraphics[angle=0,height=7cm,width=7.5cm]{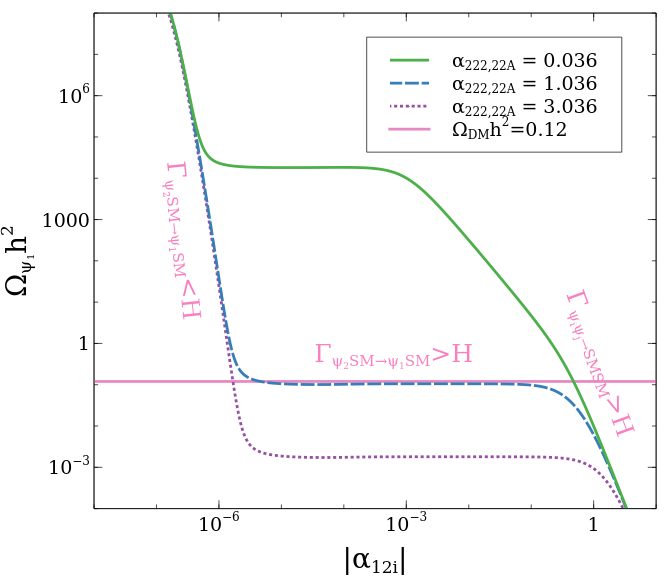}
\includegraphics[angle=0,height=7cm,width=7.5cm]{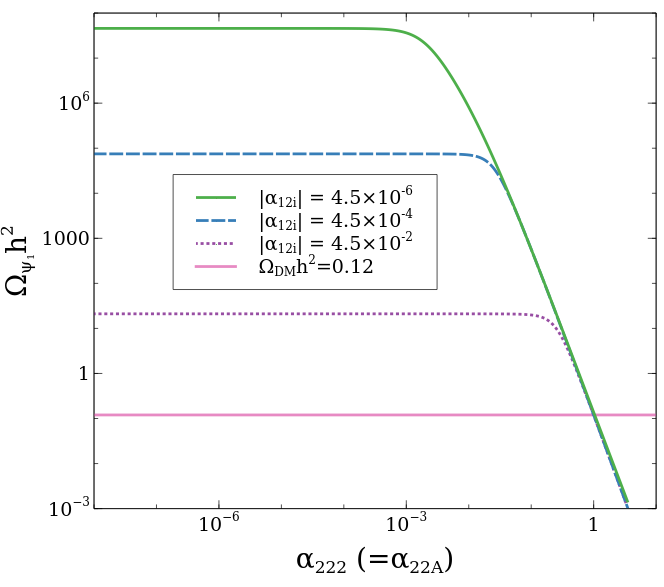}
\caption{LP (RP) shows the $\psi_1$ DM density with the variation of the 
coupling $|\alpha_{12i}|$ ($\alpha_{222}$) for three values of 
$\alpha_{222, 22A}$ ($|\alpha_{12i}|$). The other parameters 
kept fixed at values presented in the caption of Fig. \ref{line-plot-1}, and this holds
for Fig. \ref{line-plot-4} as well.} 
\label{line-plot-3}
\end{figure}

In the LP and RP of Fig. \ref{line-plot-3}, we have shown the variation 
of DM relic density with $|\alpha_{12i}|$ and $\alpha_{222,22A}$
for three values of $\alpha_{222,22A}$ and $|\alpha_{12i}|$, respectively.
In the LP, we can see DM production in three types depending on the strength
of the coupling $|\alpha_{12i}|$ (defined in Eq. \ref{psipsihi}). On the right side of the left plot, we can see DM relic density increases with the 
decrement of $|\alpha_{12i}|$ which represents DM interaction is directly 
controlled by the coupling $|\alpha_{12i}|$, then there is a plateau
which is controlled by when the freeze-out of NLSP $\psi_2$ and then a sharp increment on the left side of the plot when the process $\psi_2\, SM \rightarrow \psi_1\, SM$
goes out of equilibrium. In the figure, we can see that as we decrease the 
value of $\alpha_{222, 22A}$, we see the starting point 
of plateau shift towards the left because $\psi_2\, SM \leftrightarrow \psi_1\, SM$ process is active until the NLSP $\psi_2$ is out of equilibrium. As
we have seen from the RP of Fig. \ref{line-plot-2}, the DM density depends when
NLSP goes out of equilibrium, but it can change depending on the 
coupling strength $|\alpha_{12i}|$ as seen from the plot. In the RP of the plot, we 
have shown the DM variation with the coupling $\alpha_{222, 22A}$. We can see in the right side of the plot, DM density increases as we decrease the    
$\alpha_{222, 22A}$ couplings and after certain value it becomes constant.
This implies after certain value of $\alpha_{222,22A}$, the DM $\psi_1$
goes out of equilibrium and freezes out to a certain value. Depending on the
coupling strength of $|\alpha_{12i}|$, DM $\psi_1$ can goes out of equilibrium
in earlier times, with higher density, as clearly seen in the plot. Therefore,
from the plot, we can understand that the DM density is controlled by the 
coupling $\alpha_{222, 22A}$ up to certain point of its value after that DM
density completely controls by $|\alpha_{12i}|$ couplings value {\it i.e.}
up to certain value of $\alpha_{222, 22A}$ it is controlled by the 
$\psi_2 \psi_2 \leftrightarrow SM\, SM$ process and after that it
is controlled by the $\psi_2\, SM \leftrightarrow \psi_1\, SM$ process.

\begin{figure}[h!]
\centering
\includegraphics[angle=0,height=7cm,width=7.5cm]{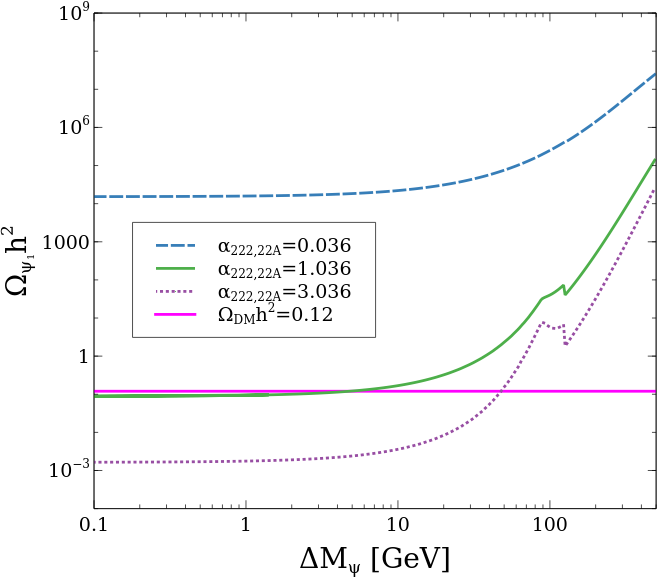}
\includegraphics[angle=0,height=7cm,width=7.5cm]{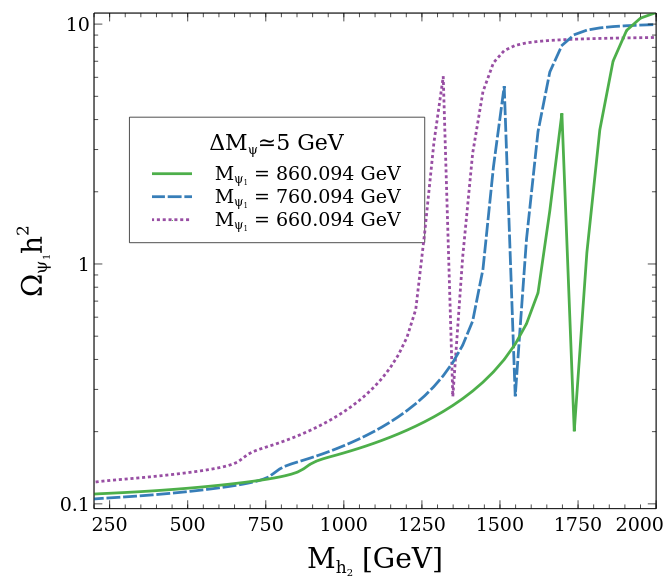}
\caption{LP and RP show the variation of DM relic density with $\Delta M_{\psi}$
and $M_{h_2}$, respectively. In the LP (RP), we have shown for three values
of $\alpha_{222, 22A}$ ($M_{\psi_1}$).} 
\label{line-plot-4}
\end{figure}

In the LP of Fig. \ref{line-plot-4}, we have shown the DM variation with the
mass difference of NLSP and DM masses for three values of $\alpha_{222, 22A}$
couplings. We can see that due to the different values of the couplings 
$\alpha_{222, 22A}$, we have different amounts of DM productions as discussed in the 
earlier plots. As we see from the plot, the DM relic density does not change
up to $\Delta M_{\psi} = 10$ GeV. This is because
DM relic density depends on the mass difference as 
$\Omega_{\psi_1}h^{2} \propto e^{\frac{\Delta M_{\psi}}{M_{\psi_2}} z_{f} }$ 
(where $20 \leq z_{f} \leq 30$ is the 
freeze-out time), so for small mass difference the effect is small, but as we increase 
the mass difference, the effect getting large and can overproduce the DM 
if we go $\Delta M_{\psi} \geq 50$ GeV (limits depend on the coupling 
$\alpha_{222, 22A}$ as well). In the RP of the figure, we have shown the 
variation of the DM relic density with BSM Higgs mass $M_{h_2}$. We can see 
as the mass of $M_{h_2}$ increases, we have enhancement in the DM density
as $M_{h_2} > M_{\psi_2}$ happens and opening of the forbidden channel of
NLSP annihilation $\psi_2 \psi_2 \rightarrow h_{2} h_{2}$ and after enhancement it has the deep due to the $h_2$ resonance. We see three resonances due to three choices of DM mass as listed in the plot.

\subsection{Scatter Plots}

In this section, we have presented scatter plots in different planes consisting of different model parameters or combinations of model parameters after obeying the constraints displayed in section \ref{constraints}. In this context, we have varied the model parameters within the following ranges,
 \begin{eqnarray}
&& 10^{-3} \leq \theta_{12,13,23} \leq 0.1, 10^{2} \leq M_{\psi_1}~[GeV] \leq 10^{3}, 1 \leq \biggl( M_{\psi_2} - M_{\psi_1} \biggr)~[GeV] \leq 100,
 \nonumber \\
&& 1 \leq \biggl( M_{h_2}-M_{h_1} \biggr)~[GeV] \leq 10^{3},
1 \leq \biggl( M_{A} - M_{Z_{BL}} \biggr)~[GeV] \leq 10^{3},
10^{-1} \leq \theta_L \leq 10^{-5},\nonumber \\
&& 0.1 \leq M_{Z_{BL}}~[GeV] \leq 10, 10^5\, (1) \leq v_{2}\, (v_{1})~[GeV] \leq 10^9\, (10^{3}),
1 \leq \biggl(M_{h_3}-M_{A} \biggr)~[GeV] \leq 10^{3}. \nonumber \\  
 \end{eqnarray}
In the above mass ranges, we observe that $M_{A} > M_{Z_{BL}}$, ensuring that the decay mode is always open for $A$ and it is not stable particle, 
and $M_{h_3} > M_{A}$ to maintain the quartic coupling $\lambda_2 > 0$. Moreover, we have also required that $M_{h_3} > 2 M_{\psi_2}$ and $M_{\psi_2} > M_{\psi_1} + M_{Z_{BL}}$ during the parameter scan, so that $\psi_2$ DM can always be produced from the resonance decay of $h_3$ at the collider, followed by the subsequent decay of $\psi_2$ to $\psi_1 X$ (X = $Z_{BL}, A, h_{1,2,3}$), preferably via a two-body decay 
involving the gauge boson $Z_{BL}$ and the CP-odd particle $A$ or three body decay
for higher values of $|\alpha_{12i}|$.
Additionally, we have chosen the vevs $v_{1,2}$ in different orders, 
ensuring that $\psi_1$ interacts very weakly with the visible sector, 
while $\psi_2$ interacts more strongly, similar to a WIMP-like DM candidate\footnote{A more hierarchical case of 
$v_{1}$  and $v_{2}$  with 
$|\alpha_{12i}| = 0$ has been studied in Ref. \cite{Khan:2025yko} in the context of a combined WIMP–FIMP scenario.}. 
This specific choice distinguishes our study from previous works and 
is being explored for the first time in the context of the present particle setup.
It is important to note that, in generating the scatter plots, 
we have followed all the constraints listed in section \ref{constraints}.

\begin{figure}[h!]
\centering
\includegraphics[angle=0,height=7cm,width=7.5cm]{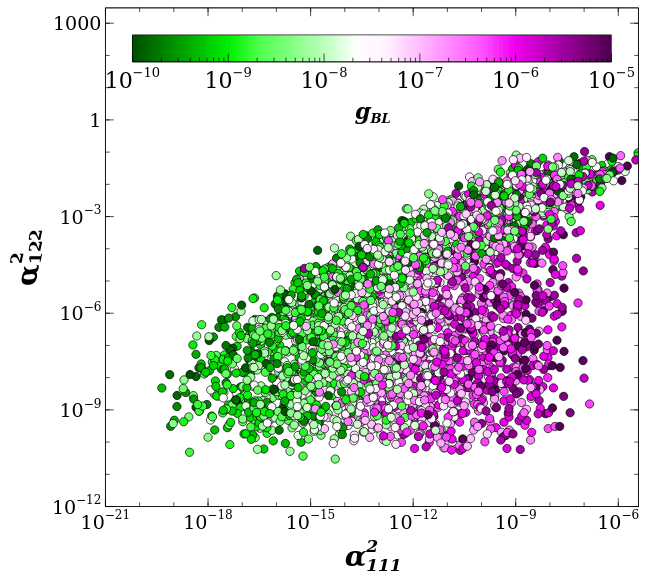}
\includegraphics[angle=0,height=7cm,width=7.5cm]{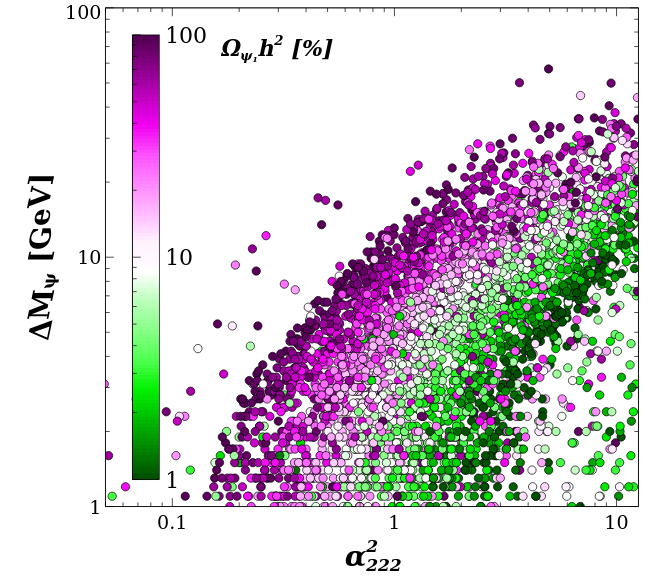}
\caption{The LP and RP show the scatter plots in the $\alpha^2_{111} - \alpha^2_{122}$ and $\alpha_{222} - \Delta M_{\psi}$ planes, respectively. In the LP, the colour bar indicates different values of the gauge coupling $g_{BL}$, whereas in the RP, it represents the percentage contribution of $\psi_{1}$ DM to the total DM density.} 
\label{scatter-plot-1}
\end{figure}

In Fig. \ref{scatter-plot-1}, we have shown scatter plots in the
$\alpha^2_{111}-\alpha^2_{122}$ and $\alpha_{222}-\Delta M_{\psi}$
planes, where the colour variation shows the different values of
gauge coupling $g_{BL}$ and the percentage of DM density compared to the total
DM density in the LP and RP, respectively. The parameters $\alpha_{111}$,
$\alpha_{122}$, and $\alpha_{222}$ have been defined in section \ref{model},
and $\Delta M_{\psi} = M_{\psi_2} - M_{\psi_1}$ is the mass difference between NLSP and DM.
In the LP, we can see the allowed region after demanding all the constraints, mainly the DM relic density bound described in section \ref{constraints}.
In the LP, we can see that $\alpha^2_{111}$ is very suppressed because
it changes with the vev $v_2$ inversely, {\it i.e.},
$\alpha_{111} \propto 1/v_2$, and the Higgs mixing matrix
off-diagonal term $U_{31}$. The colour variation also implies
that $\alpha^2_{111}$ increases with the increment of gauge coupling
$g_{BL}$, {\it i.e.}, equivalent to the decreasing value of the vev $v_2$.
Moreover, on the y-axis, $\alpha^2_{122}$ does not increase linearly
with the gauge coupling $g_{BL}$. This is because $\alpha_{122} \propto
1/v_{1}$ and $v_{1} \ll v_{2}$, so variation of $v_1$ does not change
$g_{BL}$ significantly. An important point to take from the plot is that
we can achieve DM relic density through the freeze-out mechanism even if
the DM interaction with the Higgses is in the feeble regime.
This is only possible because of the $\psi_2\, \text{SM} \rightarrow \psi_{1}\, \text{SM}$
process, which is mainly controlled by the $\alpha_{12i}, \alpha_{21i}$
($i = 1,2,3,A$). In the RP, we have shown a scatter plot in the
$\alpha_{222}-\Delta M_{\psi}$ plane. In understanding the plot, the LP and
RP of Fig. \ref{line-plot-2} will help us significantly. For a fixed
value of $\alpha_{222}$, if we go along the y-axis, we can see that the DM relic
density increases because of early decoupling of $\psi_1$ from the
thermal bath due to the suppression factor $e^{-\Delta M_{\psi}/T}$.
This behaviour is also reversed in the co-annihilation process
of DM annihilation, where DM relic density decreases with the
increment of $\Delta M_{\psi}$ \cite{Griest:1990kh}. On the other hand, by keeping a fixed value
of $\Delta M_{\psi}$ and going along the x-axis with increasing values of
$\alpha_{222}$, we see DM density decreases. This is again due to the late
freeze-out of $\psi_2$ because of the higher value of $\alpha_{222}$,
and hence the late freeze-out of $\psi_1$ as well. Moreover, for higher values of
$\alpha^2_{122}$, we can see the violation of this correlation because for
those values, $\psi_1$ can independently set its abundance without relying on $\psi_2$.
Finally, the plot implies that
we cannot choose $\Delta M_{\psi}$ and $\alpha_{222}$ arbitrarily; otherwise,
DM will be under-abundant or overabundant, which can be understood from the
colour variation in the plot.

\begin{figure}[h!]
\centering
\includegraphics[angle=0,height=7cm,width=7.5cm]{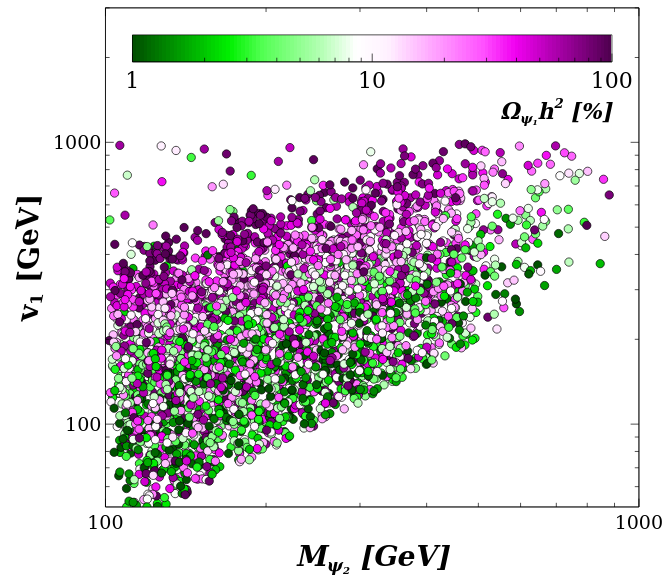}
\includegraphics[angle=0,height=7cm,width=7.5cm]{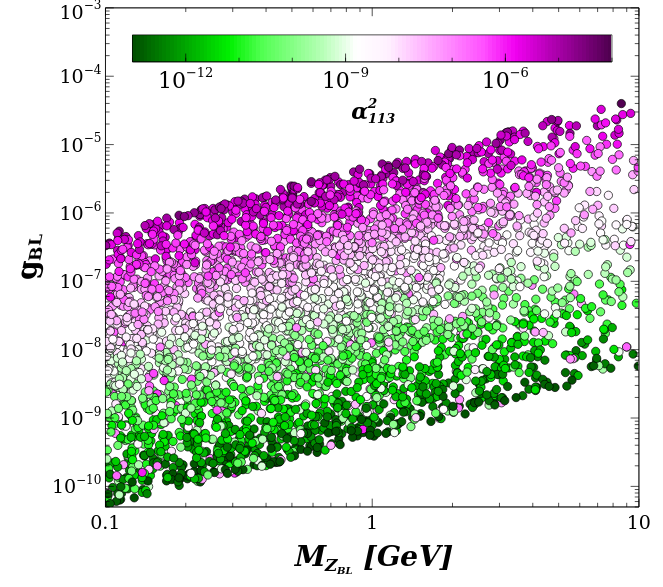}
\caption{LP shows the scatter plot in the $M_{\psi_1}-v_{1}$ plane after satisfying all the constraints, where the colour bar shows the
percentage of $\psi_1$ DM. In the RP, we have shown the
$M_{Z_{BL}}-g_{BL}$ plane, where the colour variation shows the different
values of $\alpha^2_{113}$, which is the interaction of $\psi_1$
DM with the Higgs $h_3$.} 
\label{scatter-plot-2}
\end{figure}

In the LP and RP of Fig. \ref{scatter-plot-2}, we have shown the
scatter plots in the $M_{\psi_2}-v_{1}$ and $M_{Z_{BL}}-g_{BL}$
planes, respectively. The colour variation in the LP shows the
percentage of the DM compared to the total amount, whereas the
RP represents the coupling value $\alpha^2_{113}$.
In the LP, we can see that a sharply correlated region is allowed in the $M_{\psi_2} - v_{1}$
plane after satisfying all the constraints. In particular, the lower region is
disallowed because we have demanded $|\alpha_{22i}| < \sqrt{4 \pi}$, which
implies we cannot take a very high value of $M_{\psi_2}$ for a particular value
of $v_1$. We can see that for a fixed value of $M_{\psi_2}$, if we increase the
$v_1$ value, then we decrease the coupling $\alpha_{22i}$ 
(as seen from Eq. \ref{psipsihi}),
hence the increment in the DM density (clearly visible from the colour variation).
We can also see from the colour variation that some points do not follow this trend, which arises from another effect due to the mass difference between
$\psi_2$ and $\psi_1$, as we discussed in the previous plot, and a partial effect of
the vev $v_2$ when it is around the same value as $v_1$. It is to be noted that we
have fewer dense points above $M_{\psi_2} > 500$ GeV. This is mainly
due to the requirement $M_{h_3} > 2 M_{\psi_2}$, and we have checked that higher 
DM mass regimes are also equally probable from DM physics if we omit this bound. However, those higher mass regime points will have very low sensitivity at 
the MATHUSLA or FASER detectors
at the LHC collider due to the low production cross-section for the high
value of the BSM Higgs mass. 
In the RP, we have shown the
scatter plot in the $M_{Z_{BL}}-g_{BL}$ plane. The allowed region mainly comes
from the ranges of $M_{Z_{BL}}$ and $v_2$ considered in this work. We can
see that for a fixed value of $M_{Z_{BL}}$, if we increase the gauge coupling
$g_{BL}$, then we decrease the vev $v_2$, which results in larger
values of $\alpha^2_{113}$, as seen from the colour variation very clearly.
The opposite effect is observed when we go along the x-axis for a fixed value
of $g_{BL}$.

\begin{figure}[h!]
\centering
\includegraphics[angle=0,height=7cm,width=7.5cm]{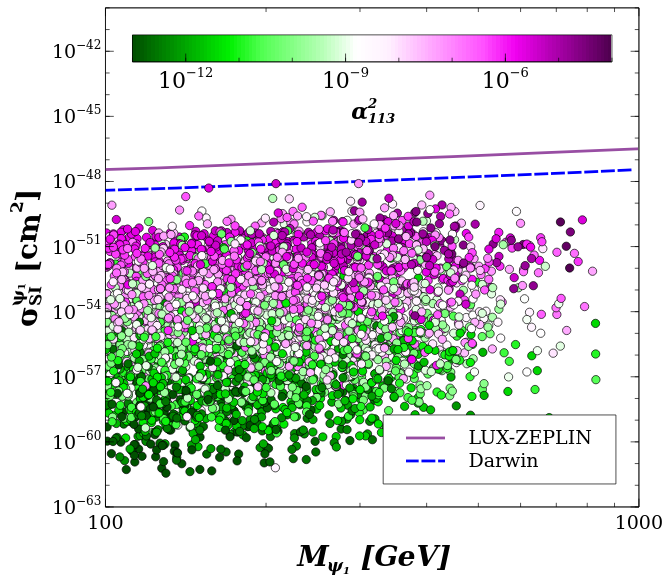}
\includegraphics[angle=0,height=7cm,width=7.5cm]{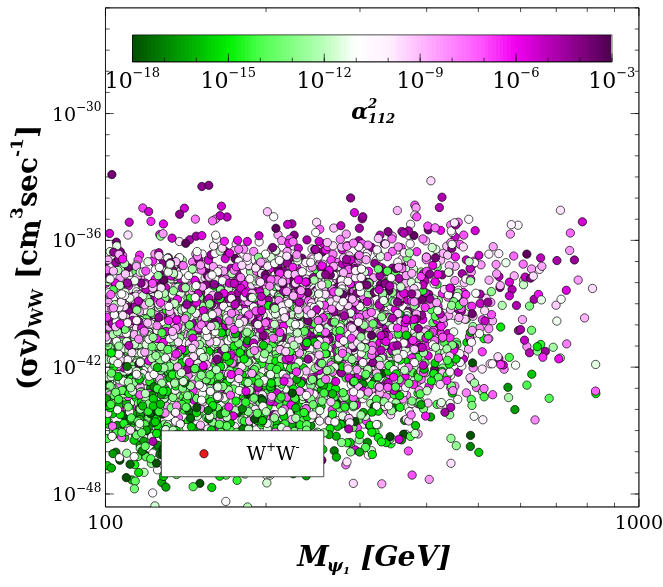}
\caption{Scatter plots in the $M_{\psi_1} - \sigma^{\psi_1}_{SI}$
and $M_{\psi_1} - (\sigma v)_{WW}$ planes have been shown, where the
colour bar in the LP (RP) shows the different values
of $\alpha^2_{113}$ ($\alpha^2_{112}$).} 
\label{scatter-plot-3}
\end{figure}

In the LP and RP of Fig. \ref{scatter-plot-3}, we have shown the scatter plots
in the $M_{\psi_1}-\sigma^{\psi_1}_{SI}$ and $M_{\psi_1} - (\sigma v)_{WW}$
planes. The colour variation in the LP and RP shows the different values
of DM coupling with $h_3$ and $h_2$, {\it i.e.}, $\alpha^2_{113}$
and $\alpha^2_{112}$, respectively. Again, in both plots, we do not have
any points for $M_{\psi_1} > 900$ GeV, mainly because of the requirement
$M_{h_3} > 2 M_{\psi_2}$. In the LP, we can see all the points lie below the
current LUX-ZEPLIN direct detection bound, even though DM is produced by the
freeze-out mechanism. In particular, the coupling responsible for
DM direct detection does not control the production of $\psi_1$ DM,
hence we can freely choose the parameter and explain the non-detection of
DM in the DD experiments so far. On the other hand, in the RP, we have shown the
context of DM indirect detection experiments. We can see that the cross-section
is sufficiently below the thermal cross-section value
$(\sigma v)_{WW} \sim 3\times 10^{-26}$ $cm^{3}sec^{-1}$ and 
much below the future projected sensitivity of indirect detection 
experiments like Fermi-LAT \cite{LSSTDarkMatterGroup:2019mwo} 
and CTA \cite{CTA:2020qlo}. The suppression
comes from two factors: one is the choice of a small Higgs portal coupling of DM
with the visible sector, and the other is p-wave suppression by the square of the DM
velocity due to fermionic DM annihilation. 
Therefore, although DM is produced by the freeze-out mechanism, it is still
below the sensitivity of current direct detection experiments and much
below that of indirect detection experiments. This mode of DM production
can naturally explain the non-detection of DM in various experiments
so far, while still offering extensive detection prospects in the near future.
At the same time, it is not completely detection-blind like the freeze-in
type of DM.

\section{Long Lived Particle signature at MATHUSLA}
\label{mathusla-section}
In the present work, we have the possibility of detecting DM at the 
MATHUSLA \cite{Curtin:2018mvb} or FASER \cite{FASER:2022hcn} detector at LHC. 
In the present work, we provide the 
prediction of our work at the MATHUSLA detector, and a similar kind of conclusion will also hold for the ongoing FASER detector. In Ref. \cite{Curtin:2018mvb}, the authors have studied the prospects 
of the MATHUSLA detector by considering different detector sizes. In the present work, we use the 14 TeV HL-LHC collider with the integrated luminosity 
$\mathcal{L}= 3 ab^{-1}$ and the detector size of
$[200mt\times 200mt\times 20 mt]$. The basic strategy will be to produce the BSM
Higgs at the collider, which promptly decays to long lived particle and then 
long lived particle decays outside the collider detectors, containing some visible 
particles. In our work, we have considered $h_3$ production at the collider, which has a suppressed cross section compared to the SM Higgs by a factor $U^2_{13}$ and then $h_{3} \rightarrow \psi_2 \psi_2$ decays happen promptly insider the 
collider detector. Finally, the produced $\psi_2$ can decay promptly or as long lived 
particle outside the detector, depending on the final state particles and mass spectrum.
Following Ref. \cite{Curtin:2018mvb}, we need to calculate the boosted decay length 
$\bar b c \tau$, where $\bar b = \frac{1.5 M_{h_3}}{2 M_{\psi_2}}$ 
\cite{Curtin:2018mvb}, 
of NLSP particle $\psi_2$. The production cross section of the long-lived 
particle, we have determined as follows,
\begin{eqnarray}
\sigma_{pp \rightarrow \psi_{2} \psi_{2} jj} = \sigma_{pp \rightarrow h_{3}} \times 2Br(h_{3} \rightarrow \psi_{2} \psi_{2}),
\end{eqnarray}
the factor 2 comes for the double $\psi_2$ production from the decay of BSM Higgs $h_3$. The production cross section we have matched with the LHC data
with proper K-factor and is taken from Ref. \cite{Covi:2025erx}.  
\begin{figure}[h!]
\centering
\includegraphics[angle=0,height=5.8cm,width=7.3cm]{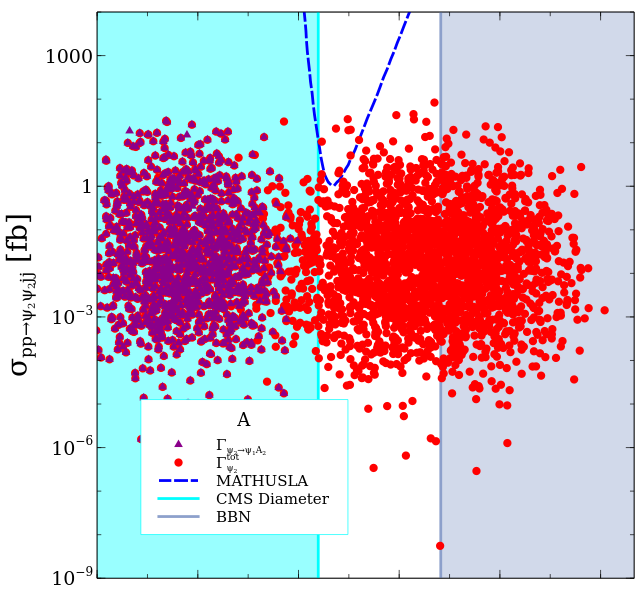}
\includegraphics[angle=0,height=5.8cm,width=7.3cm]{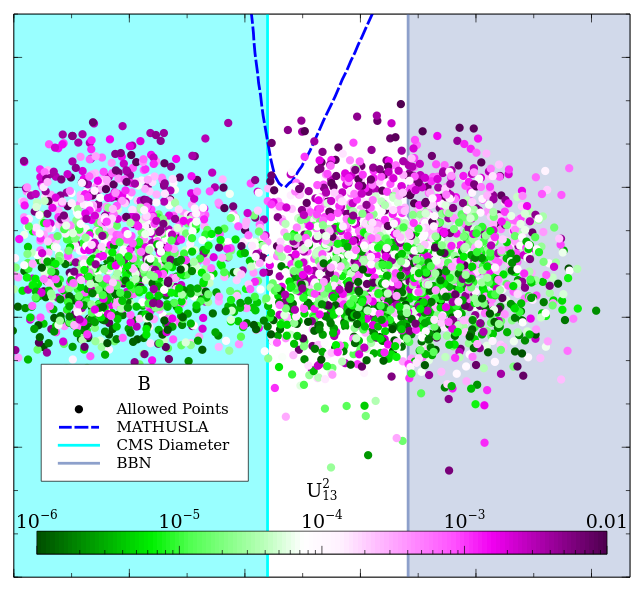}
\includegraphics[angle=0,height=5.8cm,width=7.3cm]{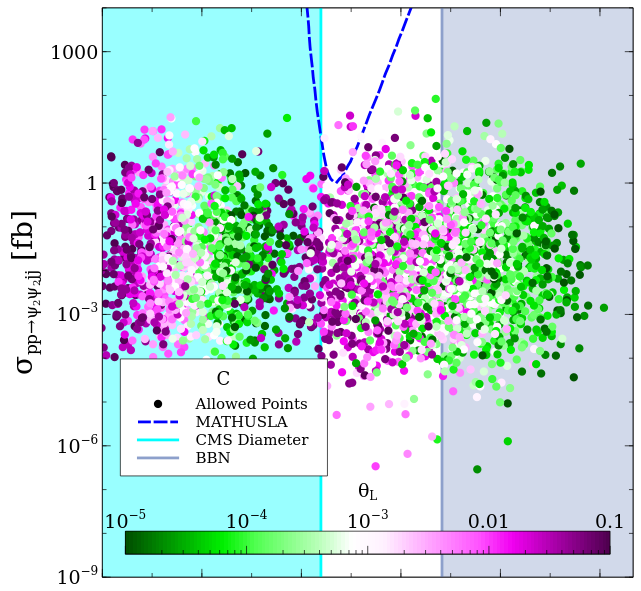}
\includegraphics[angle=0,height=5.8cm,width=7.3cm]{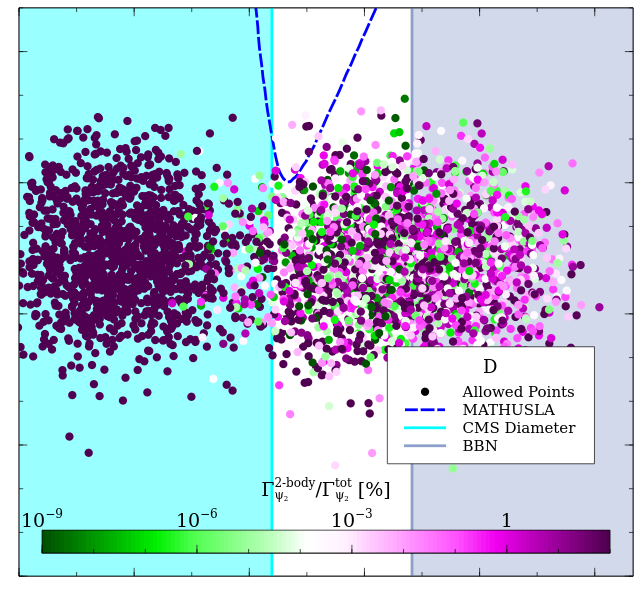}
\includegraphics[angle=0,height=5.8cm,width=7.3cm]{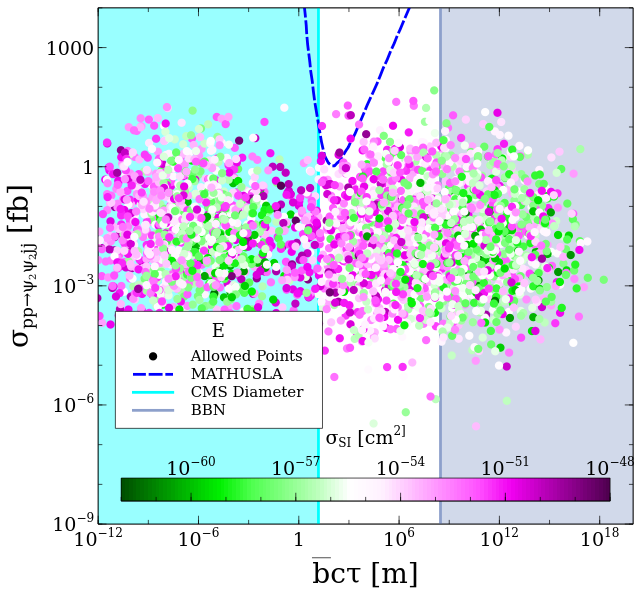}
\includegraphics[angle=0,height=5.8cm,width=7.3cm]{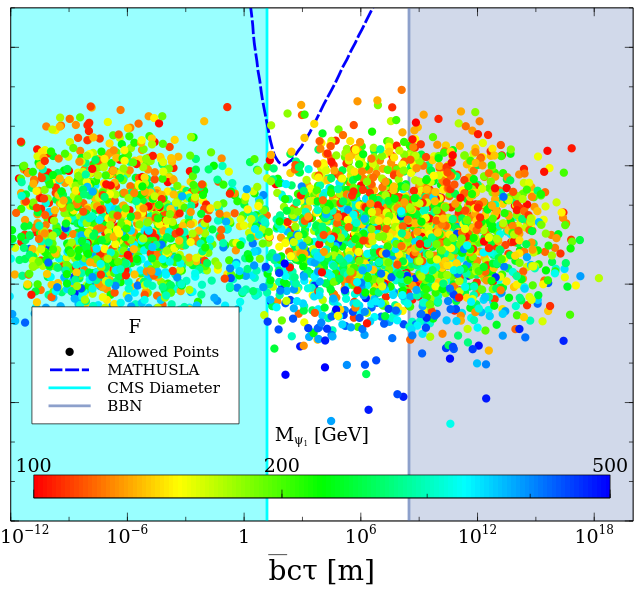}
\caption{All the plots are displayed in the $\bar{b}c\tau - \sigma_{pp \rightarrow \psi_1 \psi_1 jj}$ plane, where the colour bar in each plot represents 
different values of the model parameters as follows: $U^2_{13}$ (B), $g_{BL} \sin^{2}\theta_L$ (C), $\Gamma^{2-body}_{\psi_2}/\Gamma^{tot}_{\psi_2}$ (D),
$\sigma_{SI}$ (E), and $M_{\psi_1}$ (F). The letters in the bracket label the
corresponding figures.
The cyan region corresponds to the diameter of the CMS detector, while the grey region denotes decay lengths that exceed the BBN time.} 
\label{mathusla-plot}
\end{figure}

In Fig. \ref{mathusla-plot}, we have shown the scatter plots in the
$\bar b c \tau - \sigma_{pp \rightarrow \psi_2 \psi_2 jj}$ plane after
satisfying all the constraints listed in Section \ref{constraints}.
In Fig. \ref{mathusla-plot}-A, we have shown the magenta and red points,
which imply NLSP $\psi_2$ dominantly decays to $\psi_{1}A$ (magenta) and
$(\psi_{1} Z_{BL}+\psi_{1}f\bar f)$ (red). The expressions of different decay 
modes of $\psi_2$ for two-body and three-body are given in Eqs. (\ref{psi2-psi1ZBL}, 
\ref{psi2-psi1A}, \ref{psi2-psi1ff}). The dependence of the decay widths
depends on the parameter $\alpha_{12A}$ for $\psi_1 A$ mode, $g_{BL}$
for $\psi_1 Z_{BL}$ mode and $\alpha_{12i}$ and $U_{1i}$ ($i = 1,2,3$) for three
body decay $\psi_{1} ff$ mode. 
In the plot, the blue dashed line represents
MATHUSLA sensitivity for 14 TeV HL-LHC run with 3 $ab^{-1}$ data and the detector
size of [$200mt\times 200mt\times 20mt$]. The cyan region represents the CMS diameter,
therefore, if the $\psi_2$ decay length lies within this diameter, we can expect some
signal at the CMS detector. The grey region represents the BBN time, so the NLSP decays after
BBN and CMB may potentially alter the BBN and CMB predictions. 
All the coloured regions
imply the same meaning for the rest of the plots as well.
We can see from the plot that some part of the allowed region
can be explored at the MATHUSLA detector\footnote{The production cross-section can be increased by a factor of four because we have considered mixing angle maximally up to 0.1 which can be taken up to 0.2 after obeying the current collider bounds.}. Moreover, when $\psi_2$
dominantly decays to $\psi_1 A$, we expect some signal inside the CMS
detector, and when $\psi_2$ decays to $(\psi_1 Z_{BL} + \psi_1 ff)$, 
the decay length can
exceed the BBN bound. The exact number of points ruled out by the BBN
bound depends on the NLSP abundance, which is smaller compared to the
DM abundance. Therefore, we expect that the BBN bound will allow most of the 
points, and we have shown it in section \ref{bbn-section}.
We have checked the decay width output from micrOMEGAs using the routine
``pWidth(pName,\&address)'' and the analytical expressions in the Appendix 
in Eqs. (\ref{psi2-psi1ZBL}, 
\ref{psi2-psi1A}) and they are matching each other.
In Fig. \ref{mathusla-plot}-B, we have shown the colour bar for $U^2_{13}$,
which is related to the production cross section of $h_3$. We see a
linear variation in the colour with the production cross-section of
NLSP $\sigma_{pp \rightarrow \psi_{2} \psi_{2} jj}$. Moreover, we also
see some magenta points representing lower production cross sections,
which can correspond to a higher value of $h_3$ mass or small branching of
$h_{3}$ to $\psi_2 \psi_2$.
In Fig. \ref{mathusla-plot}-C, we have shown the
colour bar for the BSM fermion mixing angle $\theta_L$. We can see an oscillatory
pattern in the colour variation. This is because the points within the cyan region
mainly represent $\psi_{2} \rightarrow \psi_{1} A$, and the outside
points represent $\psi_{2} \rightarrow (\psi_{1} Z_{BL} + \psi_1 f \bar f)$. 
Now both decay
channels are proportional to $\sin^{2}\theta_{L}$, therefore, we see colour variation
in two regions.
In Fig. \ref{mathusla-plot}-D, the colour variation shows the ratio of the 
2-body decay of $\psi_2$ to the total decay width of $\psi_2$, including both 2-body and 3-body decays, {\it i.e.}, $\Gamma^{2-body}_{\psi_2}/\Gamma^{tot}_{\psi_2}$. 
 The values from the colour bar indicate that for the chosen model parameters, the 2-body decay is always dominant when the channel $\psi_2 \rightarrow \psi_1 A$ is open. Beyond that, either the 2-body or 3-body decay modes can be dominant, depending on the relative strengths of the parameters $\alpha_{12i}$, $U_{1i}$ ($i=1,2,3$), $g_{Bl}$, and $\theta_L$. 
Therefore, beyond the dark magenta points, we observe a scattered region with random colour variation. It is also worth noting that the associated production of the gauge boson $Z_{BL}$ results in a prompt decay with a very short lifetime. We have
 verified its decay width using both the analytical expression in 
 Eq. \ref{ZBL-decay-width} and micrOMEGAs, and the outputs are the same.
In Fig. \ref{mathusla-plot}-E, we have shown the spin-independent direct detection
cross section in the colour bar. We can see that the region inside the CMS detector
has a lower SIDD as the decay length increases. This is due to the vev $v_2$,
which controls both the DM interaction with the visible sector through the Higgses and
the decay width $\Gamma_{\psi_{2} \rightarrow \psi_{1} A}$. 
On the other hand, the regions outside
are influenced by $g_{BL}$, which affects $\bar b c \tau$ inversely and DM interaction linearly.
In Fig. \ref{mathusla-plot}-F, we have shown the
colour variation in terms of the DM mass. We can see that lower DM mass
corresponds to higher production, and higher mass corresponds to lower production.
This is again related to the mass ordering $h_{3} \rightarrow \psi_{2} \psi_{2}$
and $\psi_{2} \rightarrow \psi_{1} X$. Therefore, as we take a larger value
of $M_{\psi_1}$, it results in higher values of $M_{\psi_2}$ and
hence $M_{h_3}$. Therefore, we observe a higher production cross-section for lower masses and vice versa.
 
\section{Allowed region from BBN}
\label{bbn-section}

In this section, we aim to constrain the model parameters that yield a lifetime 
for the NLSP exceeding the BBN timescale. In the decay width of $\psi_2$, we 
deliberately focus on the region where it decays to $\psi_1 Z_{BL}$ and $\psi_1 f \bar f$, since the other decay mode, $\psi_1 A$, has no impact on BBN physics.
\begin{figure}[h!]
\centering
\includegraphics[angle=0,height=7cm,width=7.5cm]{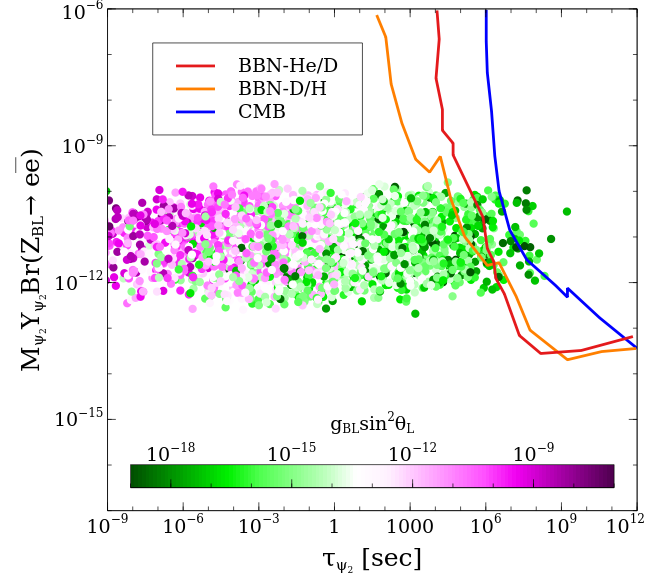}
\includegraphics[angle=0,height=7cm,width=7.5cm]{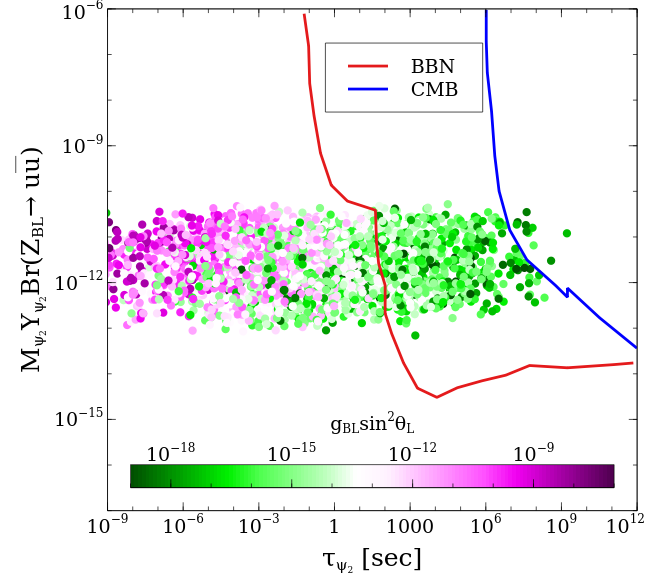}
\caption{LP and RP show the allowed regions after implementing the BBN and CMB
bounds, taking into account the NLSP abundance. We have focused on the region
where $\psi_2$ dominantly decays to $\psi_1 Z_{BL}$ and $\psi_1 f \bar f$. 
The colour bar in both
plots shows the value of $g_{BL} \sin^{2}\theta_L$.} 
\label{BBN-plot}
\end{figure}

In Fig. \ref{BBN-plot}, we have shown the scatter plots in the
$\tau_{\psi_2}-M_{\psi_2} Y_{\psi_2} Br(Z_{BL} \rightarrow e \bar{e})$
and $\tau_{\psi_2}-M_{\psi_2} Y_{\psi_2} Br(Z_{BL} \rightarrow u \bar{u})$
planes. The colour bar shows the values of
$g_{BL} \sin^{2}\theta_{L}$. We have focused on the parameter region where
$\psi_2$ decays to $\psi_1 Z_{BL}$ and $\psi_1 f \bar f$ because the other 
decay mode $\psi_{1} A$
occurs much earlier than the BBN time for the choice of our model
parameter values, as seen in Fig. \ref{mathusla-plot}-A.
In determining DM abundance, 
we have solved the coupled Boltzmann equations up to 1 MeV and taken the values 
of $\psi_1$ and $\psi_2$ abundance at that scale, and the freeze-out of DM completely depends on the NLSP freeze-out time.
During the equilibration period, the rate of change from DM to NLSP
can be expressed as,
\begin{eqnarray}
\Gamma_{\psi_{1} \rightarrow \psi_2} = \frac{n^{eq}_{\psi_2}}{n^{eq}_{\psi_1}}
\sum_{k = SM} \langle \sigma v \rangle_{\psi_{2} k \rightarrow \psi_{1} k} n^{eq}_{k}.
\end{eqnarray}
Therefore, when $\psi_2$ goes out of equilibrium, $\psi_1$ also goes out of equilibrium. Using the abundance of $\psi_1$, we can estimate the
comoving number density of $\psi_2$ approximately using the relation\footnote{For proper determination, we need to evolve the coupled Boltzmann
equations in micrOMEGAs for a lower value of the global variable ``Tend", which is just
before the decay of $\psi_2$. We expect the same output as derived using the
$\psi_1$ density, with slight deviation when $\psi_1$ and $\psi_2$ interact
strongly.}, $Y_{\psi_2} \simeq
e^{- \Delta M_{\psi}/T_{f}} Y_{\psi_1}$, where we have
defined earlier $\Delta M_{\psi} = M_{\psi_2} - M_{\psi_1}$ 
and $T_{f} = M_{\psi_2}/25$.
In the LP, we have shown the points where $Z_{BL}$ decays to $e^{+}e^{-}$,
and the bounds are shown for the He/D (red line) and D/H (brick line) 
ratios coming from BBN
and the CMB bound (blue line). The bounds are taken from Ref. \cite{Kawasaki:2017bqm},
 and similar studies in different particle spectra can be found in Ref. \cite{Belanger:2022gqc}.
We can see from the colour bar that we cannot take $g_{BL}$ and $\sin\theta_L$
arbitrarily small, and $g_{BL} \sin^{2}\theta_{L} < 10^{-14}$ would conflict with the
BBN bound. On the other hand, in the RP, we have shown the $u\bar{u}$
decay mode of $Z_{BL}$, and from the BBN bound and the colour bar, we can
conclude that $g_{BL} \sin^{2}\theta_L  < 10^{-14}$ may alter the BBN prediction.
We can see that the $u\bar{u}$ final state imposes a stronger bound than $ee$, and
in particular, we can avoid the $u\bar{u}$ bound by choosing the $Z_{BL}$ mass
below 100 MeV so that it has no hadronic decay mode open.

\section{Conclusions}
\label{conclusion}
In the present work, we have considered an alternative $U(1)_{B-L}$ model by extending the SM with four chiral fermions and two singlet scalars. The model has previously 
been studied for WIMP, FIMP, and multicomponent DM scenarios, and in the 
present work, we have, for the first time, realised the conversion-driven 
freeze-out mechanism. To accomplish CDFO, we have utilised $\psi_2$ as the 
NLSP and $\psi_1$ as the DM candidate.
The abundance of $\psi_1$ DM is entirely determined by the abundance of the NLSP and not by the interaction of DM with the visible sector. In this way, 
the direct interaction of $\psi_1$ with the visible sector can be set to any 
value without impacting the DM density. To study CDFO, we have focused on the 
parameter range where $\psi_2 \psi_2 \leftrightarrow SM~SM$ and 
$\psi_2~SM \leftrightarrow \psi_1~SM$ processes are greater than the Hubble parameter in the early Universe. Additionally, the process $\psi_1 \psi_1 \leftrightarrow SM~SM$ remains 
below the Hubble parameter and are therefore 
ineffective in controlling the DM density\footnote{This is opposite to the freeze-in mechanism, because here the DM is in thermal equilibrium, but its direct interactions with the visible sector are below the Hubble parameter.}.
When the process $\psi_2 \psi_2 \leftrightarrow SM~SM$ goes out of equilibrium, 
$\psi_1$ abundance is fixed by the conversion process $\psi_{2}~SM \leftrightarrow \psi_{1}~SM$. After $\psi_1$ and $\psi_2$ completely decouple and 
freeze out to particular values, $\psi_2$ later decays to $\psi_1$, increasing 
the DM abundance, similar to the superWIMP mechanism.

We have found that the DM density can be easily tuned to a desired value by changing the mass difference between the NLSP and DM, as well as the NLSP interaction with the visible sector. We have also found that the mass difference between NLSP and DM cannot exceed 60 GeV, otherwise, it will overproduce the DM density, contrary to the 
co-annihilation dominated processes, where the density decreases with increasing mass 
difference \cite{Griest:1990kh}. 
We have observed that a moderate hierarchy between the vevs $v_1$ and $v_2$ can lead to change in the interactions of $\psi_2$ and $\psi_1$ with the Higgs bosons,
which can control the $\psi_2$ abundance and hence the $\psi_1$ abundance.

For the parameter range considered in this study, the direct detection cross section remains below the LUX-ZEPLIN bound but within reach of the DARWIN experiment after successful DM production via the freeze-out mechanism. Therefore, the present work naturally explains why the direct detection parameter space is yet to be fully explored by current experiments. In contrast, indirect detection experiments are 
beyond the detection prospects of the present set-up due to a double suppression: 
one from the small DM-Higgs interactions, and the second from the $p$-wave annihilation of DM, which results in a velocity-squared suppression of the cross section.
Moreover, we have also explored the detection prospects in the context of long-lived particle searches at MATHUSLA (similar conclusions holds for FASER as well). 
This scenario arises mainly from the production of the NLSP $\psi_2$ at colliders via the prompt decay of $h_3$, after which $\psi_2$ can decay either inside or outside the LHC detectors depending on whether it decays to $\psi_1 A$ or $(\psi_1 Z_{BL} + \psi_{1} f \bar{f})$, respectively. 
Finally, we have found that some parameter points can make $\psi_2$ long-lived 
up to BBN or CMB timescales, potentially altering BBN physics or the CMB anisotropies. 
We have identified the parameter values that could pose such serious problems within the standard cosmological scenario. Therefore, our DM candidate remains largely 
unexplored by current experiments, as it can naturally exhibits suppressed 
interactions, providing a compelling explanation for the lack of DM detection so far.

\section*{Acknowledgements}
SK acknowledges valuable discussions with Hyun Min Lee.
The research is supported by Brain Pool program funded by the Ministry of Science and ICT through the National Research Foundation of Korea\,(RS-2024-00407977) and Basic Science Research Program through the National Research Foundation of Korea (NRF) funded by the Ministry of Education, Science and Technology (NRF-2022R1A2C2003567).
For the numerical analysis, we have used the Scientific Compute Cluster at GWDG, the joint data centre of Max Planck Society for the Advancement of Science (MPG) and University of G\"{o}ttingen.
\appendix
\section*{Appendix: Analytical expressions for decay widths and collision terms}
\label{App:AppendixA}

\section{Higgses Mixing Matrix}

The exact form of the Higgses mixing matrix has been taken as for generating the
line plots in section \ref{line-plots-dm},
\begin{eqnarray}
U_{ij} = 
\begin{pmatrix}
0.999 & 10^{-6} & 6.37 \times 10^{-5} \\
-10^{-5} & 0.999 & 1.38\times 10^{-7} \\
-1.13 \times 10^{-6} & -1.38 \times 10^{-6} & 0.999
\end{pmatrix},
\label{higgs-mixing-matrix}
\end{eqnarray}
where the matrix is equivalent to the unit matrix.

\section{Decay widths:}
\begin{itemize}
\item The decay width of $\psi_{2} \rightarrow \psi_{1} Z_{BL}$
can be expressed as,
\begin{eqnarray}
\Gamma_{\psi_{2} \rightarrow \psi_1 Z_{BL}} &=& \frac{g^2_{BL} 
\sin^{4}\theta_{L}}{72 \pi M^2_{Z_{BL}} M_{\psi_2} } 
\biggl[ \biggl( 1 - \left(\frac{M_{\psi_1} + M_{Z_{BL}}}{M_{\psi_2}}\right)^{2} \biggr) \biggl( 1 - \left(\frac{M_{\psi_1} - M_{Z_{BL}}}{M_{\psi_2}}\right)^{2} \biggr)  \biggr]^{1/2} \nonumber \\
&\times & \biggl[ \left( M^2_{\psi_2} - M^2_{\psi_1} \right)^{2}
+ \left( M^2_{\psi_2} + M^2_{\psi_1} \right) M^2_{Z_{BL}}
- 2 M^{4}_{Z_{BL}} \biggr]
\label{psi2-psi1ZBL}
\end{eqnarray}
\item The decay width for $\psi_{2} \rightarrow \psi_{1} A$ can be
expressed as,
\begin{eqnarray}
\Gamma_{\psi_{2} \rightarrow \psi_{1} A} &=& \frac{1}{32 \pi M_{\psi_2}}
\biggl[ \left( \alpha^2_{12A} + \alpha^2_{21A} \right) 
\left(M^2_{\psi_2} - M^2_{A} + M^2_{\psi_1} \right) - 4 \alpha_{12A} 
\alpha_{21A} M_{\psi_2} M_{\psi_1} \biggr]\nonumber \\
&\times& \sqrt{\biggl[ 1 - \left( \frac{M_{\psi_1} + M_{A}}{M_{\psi_2}} \right)^{2} \biggr] \biggl[ 1 - \left( \frac{M_{\psi_1} - M_{A}}{M_{\psi_2}} \right)^{2} \biggr]} 
\label{psi2-psi1A}
\end{eqnarray}

\item The decay width of three body decay of $\psi_2$ to $\psi_1 f \bar f$
takes the following form,
\begin{eqnarray}
\Gamma^{3-body}_{\psi_{2} } = \frac{ n^c_{f}M_{\psi_2} M_{f}}{32 (2\pi)^{3} v^{2}}
\int_{x^{-}_{3}}^{x^{+}_{3}} \int_{x^{-}_{1}}^{x^{+}_{1}}\sum_{i,j=1}^{3} \frac{
U_{1i} U_{1j}
 \alpha_{12i} \alpha_{12j} \left[ 2 M_{\psi_2} \left( M^2_{\psi_1}
- 4 m^2_{f} - M^2_{\psi_2} (-1 + x_1 ) \right)   \right] dx_{3} dx_{1}}{ 
\left[M^2_{\psi_2} (-1+x_{1}) + M^2_{h_i} - M^2_{\psi_1} \right]
\left[M^2_{\psi_2} (-1+x_{1}) + M^2_{h_j} - M^2_{\psi_1} \right]  } \nonumber \\
\label{psi2-psi1ff}
\end{eqnarray}
where $n_f^c$ is the colour charge of the fermion $f$ and other quantities are defined in the main text. The integration limits can be
expressed as,
\begin{eqnarray}
x^{\pm}_{1} &=& 1 + a_{1} - a_{2} + a_{3} - x_{3} - \frac{1}{2} 
\left( 2 a_{3} - x_{3} \right)\times \left( 1 - \frac{a_{1} - a_{2}}
{1 + a_{3} - x_{3}} \right) \nonumber \\
&\pm& \frac{\sqrt{x^2_{3} - 4 a_{3}}}{2} \sqrt{1 - 2 \frac{a_{1} + a_{2}}{1 + a_{3} - x_{3}} 
+ \frac{(a_{1}-a_{2})^{2}}{(1+a_{3} - x_{3})^{2}} } \nonumber \\
x^{-}_{3} &=& 2 \sqrt{a_{3}}, x^{+}_{3} = 1 + a_{3} - a_{1} -a_{2} - 2 \sqrt{a_{1} a_{2}}
\end{eqnarray}
where $a_{i} = \left( \frac{M_{i}}{M} \right)^{2}$.

\item The decay width of $Z_{BL} \rightarrow f \bar f$ can be 
expressed as,
\begin{eqnarray}
\Gamma_{Z_{BL} \rightarrow f \bar f} = \frac{n_{2} g^2_{BL} q^2_{f}}
{12 \pi M_{Z_{BL}}} \left(2 m^2_{e} + M^2_{Z_{BL}} \right)
\biggl(1 -\frac{4 m^2_{e}}{M^2_{Z_{BL}}} \biggr)^{1/2}\,.
\label{ZBL-decay-width}
\end{eqnarray}

\end{itemize}

\end{document}